\documentclass[aps,pra,twocolumn,preprintnumbers,nofootinbib]{revtex4}
\newcommand{\beq}{\begin{equation}} 
\newcommand{\eeq}{\end{equation}} 
\newcommand{\beqa}{\begin{eqnarray}} 
\newcommand{\eeqa}{\end{eqnarray}} 
\newcommand{\beqan}{\begin{eqnarray*}} 
\newcommand{\eeqan}{\end{eqnarray*}} 
\newcommand{\ba}{\begin{array}} 
\newcommand{\ea}{\end{array}} 
\newcommand{\no}{\nonumber} 

\newcommand{\lets}{\stackrel{<}{_\sim}}

\newcommand{\ve}{\varepsilon}

\newcommand{\wt}{\widetilde} 
\newcommand{\wh}{\widehat}

\newcommand{\cO}{{\cal O}} 
\newcommand{\cP}{{\cal P}}

\newcommand{\cS}{{\cal S}}

\newcommand{\nn}{\nonumber \\}

\newcommand{\bea}{\begin{eqnarray}} 
\newcommand{\eea}{\end{eqnarray}}

\begin{document}
\preprint{UWThPh-2008-9}
\title{The $\mbox{\boldmath $K_{\ell 3}$}$ 
scalar form factors in the standard model}
\thanks{This work was supported in part by the EU Contract No. 
MRTN-CT-2006-035482, \lq\lq FLAVIAnet". }
\author{A. Kastner}
\author{H. Neufeld} 
\affiliation{Fakult\"at f\"ur Physik, Universit\"at 
Wien, Boltzmanngasse 5, A-1090 Wien, Austria}
\date{May 15, 2008}
\begin{abstract}
We discuss the predictions of the standard model for the scalar 
form factors of $K_{\ell3}$ decays. Our analysis is based on the results 
of chiral perturbation theory, large $N_c$ estimates of low-energy 
couplings and dispersive methods. It includes a 
discussion of 
isospin violating effects of strong and electromagnetic origin.
\end{abstract}
\maketitle
\section{Introduction}
\label{sec:0}
\renewcommand{\theequation}{\arabic{section}.\arabic{equation}}
\setcounter{equation}{0}

Recent high-statistics mesurements of the $K_{\ell 3}$ form factor parameters 
$\lambda_+^\prime$, $\lambda_+^{\prime \prime}$, $\lambda_0$ are available from 
ISTRA+ \cite{ISTRA}, KTeV \cite{KTeV}, NA48 \cite{NA48} and KLOE \cite{KLOE}. Only
ISTRA+ has analyzed the charged kaon decay $K^- \to \pi^0 \mu^- \nu$, whereas the 
results of the other three experiments are based on $K_L$ decays.
In particular for the scalar slope, the NA48 results are difficult to accomodate 
with those of the other experiments and the actual value of this quantity is not 
yet finally settled. A global analysis of the present experimental situation can be 
found in a recent compilation of the FLAVIAnet Kaon Working Group \cite{FLAVIAnet}.

For a comparison with the experimental outcomes, we need a theoretical prediction, 
as precise as possible, for the behaviour of the scalar form factors of $K^0_{\ell 3}$
and $K^+_{\ell 3}$ decays. We wish to address the following questions:
\begin{itemize}
\item
Which of the values of $\lambda_0$ found by the different experimental groups are 
compatible with the standard model of particle physics?
\item
Which size of isospin violation can be expected for the scalar form factors?
\end{itemize}
To this end we collect and extend the present theoretical information on the 
scalar $K^0_{\ell 3}$ and $K^+_{\ell 3}$ form factors. The principal theoretical tool 
of such an analysis is chiral perturbation theory \cite{CHPT,GL85}, the 
low-energy effective theory of the standard model. It exploits the special r\^{o}le 
of the pseudoscalars $\pi$, $\eta$, $K$ as Goldstone particles of spontaneous 
chiral symmetry breaking.
This effective quantum field theory 
can be extended by the 
inclusion of photons \cite{Urech,nr95,nr96} and leptons \cite{lept} as active 
degrees of freedom. As a complementary method, also dispersion techniques 
have been employed
\cite{JOP02,JOP04,JOP06,BOPS06,Passemar07,Bernard-Passemar07}.

The outline of this paper is as follows. In Sect. \ref{sec: basics} we recapitulate
the basic definitions of the $K_{\ell 3}$ form factors, their current parametrizations
and the Callan-Treiman theorem \cite{CT66}. We determine the quantity 
$F_K / F_\pi f_+^{K^0 \pi^-} \! (0)$
(which is one of the basic input parameters of our analysis) by combining the relevant 
theoretical expressions with recent experimental data. This section is closed with a
short review of the present theoretical and experimental status of the scalar slope 
parameter. 

In Sect. \ref{sec: form factors} we display the formulae for the 
$K_{\ell 3}$ form factors at next-to-leading order (NLO) in the chiral expansion 
including strong isospin violation end electromagnetic contributions of the orders 
$(m_d-m_u) p^2$ and $e^2 p^2$ given in \cite{CKNRT02}. For later convenience, we also 
express the $f_-$ form factors through the ratio $F_K / F_\pi$ \cite{lept} instead of 
the low-energy coupling $L_5^r$. The corresponding expressions for the scalar form
factors and slope parameters are shown in Sect. \ref{sec: Scalar form factors} and 
those relevant for the Callan-Treiman relation in Sect. \ref{sec: Callan-Treiman relations}. 
In Sect. \ref{sec: Size of isospin breaking} we update the numerical value of the 
parameter $\varepsilon^{(2)}$ which determines the size of strong isospin breaking. 
Our numerical results at NLO are then presented in Sect. \ref{sec: Numerics}.
They show, for the first time, a quantitative comparison of strong 
isospin breaking and electromagnetic contributions  generated at this chiral order. 

In the next step, we consider effects arising at NNLO. Our discussion in Sect. 
\ref{sec: p6} starts with an analysis in the limit of isospin conservation. 
We combine the two-loop results obtained in \cite{BT03} with a 
determination of the associated $p^6$ counterterms. The relevant low energy couplings 
$C_{12}^r$ and $C_{34}^r$ can be estimated by using the $1/N_C$ expansion and truncating
the hadronic spectrum to the lowest lying resonances \cite{CEEKPP05}. We show that 
the uncertainty in the determination of these coupling constants given in 
\cite{CEEKPP05} can be reduced considerably by a careful consideration of the
two-loop renormalization group equation \cite{BCE00} and by employing the analysis of
\cite{CPEN03} for the determination of the scalar and pseudoscalar resonance parameters. 
In this way, we obtain a rather accurate theoretical preditions for the slope and 
curvature parameters in the isospin limit. The analogous analysis is made for the 
size of the scalar form factor at the Callan-Treiman point and at zero momentum transfer.
In Sect. \ref{sec: Dispersive Analysis}
we compare our findings for the slope and 
the curvature of the scalar form factor with the results 
obtained by dispersive methods
\cite{JOP02,JOP04,JOP06,BOPS06,Passemar07,Bernard-Passemar07}. 

In Sect. \ref{sec: p6isobreak} we extend the results obtained at the order 
$(m_d - m_u) p^4$ \cite{Bijnens-Ghorbani07} on the $K_{\ell 3}$ scalar form 
factors by an estimate of the associated local contributions relevant for 
the splitting  
$\lambda_0^{K^0 \pi^+} \! -\lambda_0^{K^+ \pi^0}$. 
Finally, we discuss the possible size of corrections to the Callan-Treiman relation
induced by isospin violation at this chiral order.

Our conclusions are summarized in Sect. \ref{sec: Conclusions}.

\section{Basic facts}
\label{sec: basics}
\renewcommand{\theequation}{\arabic{section}.\arabic{equation}}
\setcounter{equation}{0}

The $K_{\ell 3}$ decays 
\beqa 
\label{Kplusl3}
K^+ (p_K) &\to& \pi^0(p_{\pi}) \ell^+ (p_{\ell}) \nu_{\ell}(p_{\nu}), \\
\label{K0l3}
K^0 (p_K) &\to& \pi^-(p_{\pi}) \ell^+ (p_{\ell}) \nu_{\ell}(p_{\nu}),
\eeqa
(and their charge conjugate modes) allow the exploration of the hadronic 
matrix elements
\beqa
&& \big{\langle} \pi^0(p_{\pi}) | V_{\mu}^{4-i5} (0) | K^+(p_K) 
\big{\rangle}  \\
&& = \frac{1}{\sqrt{2}} \Big[f_+^{K^+ \pi^0} (t) (p_K + p_{\pi})_\mu 
+ f_-^{K^+ \pi^0} (t) (p_K - p_\pi)_\mu \Big] , \no
\eeqa
and
\beqa
&& \big{\langle} \pi^-(p_{\pi}) | V_{\mu}^{4-i5} (0) | K^0(p_K) 
\big{\rangle}  \\
&& = f_+^{K^0 \pi^-} (t) (p_K + p_{\pi})_\mu 
+ f_-^{K^0 \pi^-} (t) (p_K - p_\pi)_\mu . \no
\eeqa
The processes (\ref{Kplusl3}) and (\ref{K0l3}) thus involve the four 
$K_{\ell 3}$ form factors $f_{\pm}^{K^+ \pi^0}(t)$, $f_{\pm}^{K^0 \pi^-}(t)$, 
which depend on 
\beq
t = (p_K - p_\pi)^2 = (p_\ell + p_\nu)^2,
\eeq
the square of the four momentum transfer to the leptons. Only $K_{\mu 3}$ 
decays are sensitive to both $f_+^{K \pi}$ and $f_-^{K \pi}$. With
$K_{e 3}$ decays only $f_+^{K \pi}$ can be tested, as in this case $f_-^{K 
\pi}$ enters together with the tiny 
quantity $m_e^2 / M_K^2 \simeq 10^{-6}$ in the formula for the Dalitz plot 
density. 

$f_+^{K \pi}$ is referred to as the vector form factor, because it 
specifies the $P$-wave projection of the crossed channel matrix elements 
$\langle 0 | V_{\mu}^{4-i5} (0) | K \pi \rangle$. The 
$S$-wave 
projection is described by the scalar form factor
\beq
\label{f0}
f_0^{K \pi}  (t) = f_+^{K \pi} (t) + \frac{t}{M_K^2 - M_\pi^2} f_-^{K 
\pi}  (t),
\eeq
which implies
\beq
f_0^{K \pi}  (0) = f_+^{K \pi}  (0) .
\eeq

Older analyses of $K_{\ell 3}$ data usually used the linear 
parametrization
\beq
f_{+,0}^{K\pi} (t) = 
f_{+}^{K\pi}  (0) \bigg( 1 + \lambda_{+,0} \frac{t}{M_{\pi^+}^2} \bigg).
\eeq
More recent high statistics experiments also 
search for a quadratic term in the form factor expansion of $f_+^{K 
\pi}(t)$, 
\beq
\label{quadpar}
f_+^{K\pi}  (t) = 
f_+^{K\pi}  (0) \bigg[1 + \lambda_+^\prime \frac{t}{M_{\pi^+}^2}
+ \frac{1}{2} \lambda_+^{\prime \prime} \big(\frac{t}{M_{\pi^+}^2} \Big)^2 
\bigg] .
\eeq
Alternatively, also a pole fit,
\beqa
\label{polefitplus}
f_{+}^{K \pi}  (t) &=& 
f_{+}^{K \pi} (0)  \frac{M_{V}^2}{M_{V}^2 -t}, \\
\label{polefit0}
f_{0}^{K \pi} (t) &=& 
f_{+}^{K \pi} (0)  \frac{M_{S}^2}{M_{S}^2 -t},
\eeqa
has been employed. Recently, a dispersive representation of the 
scalar form factor 
based on a twice subtracted dispersion relation was proposed in 
\cite{BOPS06,Passemar07,Bernard-Passemar07}:
\beqa
\label{disprep}
\frac{f_{0}^{K \pi} (t)}{f_{+}^{K \pi} (0)} &=&  
\exp\bigg[\frac{t}{\Delta_{K \pi}} \big(\ln C - G(t) \big) \bigg], 
\\
G(t) &=& \frac{\Delta_{K \pi} (\Delta_{K \pi} -t)}{\pi}
\int\limits_{t_{K \pi}}^\infty \frac{ds}{s} 
\frac{\phi(s)}{(s-\Delta_{K \pi}) (s-t-i \epsilon)}. \nonumber
\eeqa
The quantity $t_{K \pi}$ denotes the threshold of $K \pi$ scattering 
and $\Delta_{K \pi} = M_K^2 - M_{\pi}^2 $.

From the theoretical point of view, the scalar $K_{\ell 3}$ form factor 
has a remarkable property: The low-energy theorem of Callan and Treiman 
\cite{CT66} predicts the size of $f_0^{K \pi} (t)$ at the 
(unphysical) momentum transfer $t =  \Delta_{K \pi}$ 
as 
\beq
\label{Callan-Treiman}
f_{0}^{K \pi}(\Delta_{K \pi}) = \frac{F_K}{F_\pi} + \Delta_{\rm CT} , 
\eeq
with a correction term $\Delta_{\rm CT}$ of $\cO (m_u, m_d, e^2)$.
In the isospin limit ($m_u = m_d$, $e = 0$), and at first nonleading 
order, the 
tiny value $\Delta_{\rm CT} = - 3.5 \times 10^{-3}$ was worked out 
already some time ago \cite{gl852}.

Assuming for a moment a strict linear behaviour of the scalar form factor 
in the range between $t = 0$ and the Callan-Treiman point $t = \Delta_{K 
\pi}$, the slope parameter would be given by 
\beq
\label{lambda0lin}
\lambda_0 \simeq \frac{M_{\pi^+}^2}{\Delta_{K \pi}} 
\bigg( \frac{F_K}{F_\pi f_+^{K \pi} (0)} -1 \bigg)
\eeq
as a consequence of (\ref{Callan-Treiman}). The ratio
$F_K / F_\pi f_+^{K \pi} (0) $
appearing in (\ref{lambda0lin})
can be determined with remarkable precision from experimental input 
independent of $K_{\mu 3}$ data. 

Because of its central importance for our subsequent analysis, we describe 
here the determination of
\beq
\label{magrat}
\frac{F_K}{F_\pi f_+^{K^0 \pi^-} (0)}
\eeq
in some detail. We want to point out that the decay constants used here 
always refer to the respective charged pseudoscalars ($F_\pi \equiv 
F_{\pi^0}$,
$F_K \equiv F_{K^+}$). In the case of the pion, the distiction between charged 
and neutral decay constant amounts to  a tiny effect of order $(m_d - m_u)^2$, whereas
$F_{K^+}$ differs from $F_{K^0}$ by terms of order $m_d - m_u$ 
\cite{GL85}.

Including electromagnetic corrections \cite{lept,MS93}, the ratio of the 
(fully inclusive)
$K_{\ell 2 (\gamma)}$ and $\pi_{\ell 2 (\gamma)}$ widths can be written as
\beqa
\label{ratio1}
\frac{\Gamma(K_{\ell 2 (\gamma)})}{\Gamma(\pi_{\ell 2 (\gamma)})}
&=& 
\frac{|V_{us}|^2 F_K^2 M_{K^\pm} (1-z_{K \ell})^2}
{|V_{ud}|^2 F_\pi^2 M_{\pi^\pm} (1-z_{\pi \ell})^2}
\no \\
&\times& \Bigg\{1 + \frac{\alpha}{4 \pi} 
\bigg[H(z_{K \ell}) - H(z_{\pi \ell})
\no \\ 
&& {} + (3  - Z)
\ln \frac{M_K^2}{M_\pi^2} + \ldots \bigg] \Bigg\} ,
\eeqa 
where $z_{P \ell} = m_{\ell}^2 / M_P^2$. An explicit expression for the  
function $H(z)$ can be found in \cite{lept}. The chiral coupling 
\cite{lept} $Z \simeq 0.8$ arises from the electromagnetic mass difference 
of the pion,
\beq
M_{\pi^\pm}^2 - M_{\pi^0}^2 = 2 e^2 Z F_0^2 ,
\eeq
where $F_0$ denotes the pion decay constant in the chiral limit.
The dots in (\ref{ratio1}) refer to contributions arising at 
$\cO(e^2 p^4)$. Inserting the measured widths \cite{PDG06}
\beqa
\label{Kl2}
\Gamma (K_{\mu 2 (\gamma)}) &=& 0.5122(15) \times 10^8 \, {\rm s}^{-1} , 
\\
\Gamma (\pi_{\mu 2 (\gamma)}) &=& 0.38408(7) \times 10^8 \, {\rm s}^{-1} ,
\eeqa
we find
\beq
\label{ratio2}
\frac{|V_{us}| F_K}{|V_{ud}| F_\pi} = 0.27567(40)(2)(29) = 0.27567(50).
\eeq 
The first two separated errors correspond to the experimental 
uncertainties of the $K_{\mu 2 (\gamma)}$ and $\pi_{\mu 2 (\gamma)}$ 
width, respectively. The third one is an estimate\footnote{See also 
\cite{CR07} for a recent calculation of $\cO(e^2 p^4)$ contributions to 
the ratio $R^{\pi, K}_{e/\mu}$.} of the unknown 
electromagnetic contributions of 
$\cO(e^2 p^4)$. Using (\ref{ratio2}), the quantity 
(\ref{magrat}) we are interested in, can now be written as
\beq
\frac{F_K}{F_\pi f_+^{K^0 \pi^-} (0)} = 
0.27567(50) \times \frac{|V_{ud}|}{|V_{us}| f_+^{K^0 \pi^-} \! (0)} .
\eeq
For the determination of the product $|V_{us}| f_+^{K^0 \pi^-} \! (0)$, we 
employ the frequently used formula \cite{PDG06}
\beqa
\label{master formula}
\Gamma(K_{\ell 3 (\gamma)}) &=& 
\frac{G_F^2 M_K^5}{192 \pi^3} 
C_K^2 |V_{us}|^2 f_+^{K^0 \pi^-}(0)^2 I_K^\ell \no \\ 
&\times&
S_{\rm EW}
\big( 1 + \delta^{\ell}_K + \delta_{\rm SU(2)} \big) .
\eeqa
In order to avoid any bias from $K^+_{e3}$ 
(which would require additional theoretical input for the determination of 
$\delta_{\rm SU(2)}$)
or $K_{\mu 3}$ data (involving also information about $\lambda_0$, 
the quantity we actually want to determine), we are  
exclusively using input 
from $K^0_{L e 3}$ decays \cite{KTeV,Ke3}
as given in \cite{PDG06}: 

\beq
\Gamma(K^0_{L e 3 (\gamma)}) = 0.0792(4) \times 10^8 \, {\rm s}^{-1}, 
\eeq
\beq
\lambda_+^\prime = 0.0249(13) , \, \,
\lambda_+^{\prime \prime} = 0.0016(5) , \, \,
\rho_{\lambda^\prime, \lambda^{\prime \prime}} \simeq - 0.95 .
\eeq
Taking into account the recently determined values \cite{Moussallam05}
of the electromagnetic low energy couplings $X_i$ \cite{lept}, we obtain 
\beq
\delta_{K^0}^e = 0.0114(30)
\eeq
as an update of our electromagnetic corrections presented in \cite{CNP04}.
Putting everything together, we find
\beq
\label{magprod}
|V_{us}| f_+^{K^0 \pi^-} \! (0) = 0.21616(68) .
\eeq
With  \cite{TH07}
\beq
\label{Vud}
|V_{ud}| = 0.97418(26) ,
\eeq
extracted from superallowed nuclear Fermi transitions, we finally 
obtain\footnote{The small difference between our number and the one 
obtained in \cite{JOP06} within a similar approach is due to the 
slightly different input parameters.} 
\beq
\label{valmagrat}
\frac{F_K}{F_\pi f_+^{K^0 \pi^-} \! (0)} =  1.2424(23)(39)(3) = 
1.2424(45),
\eeq
where the first error comes from (\ref{ratio2}), the second one from 
(\ref{magprod}) and the third one from (\ref{Vud}).

Inserting (\ref{valmagrat}) in (\ref{lambda0lin}) gives $\lambda_0 \simeq 
0.02$ as a rough estimate. In reality, as it is also suggested by the 
pole parametrization (\ref{polefit0}), the second derivative of the scalar 
form factor is positive in the physical region and 
\beq
\lambda_0^{K \pi} := \frac{M_{\pi^+}^2}{f_+^{K \pi} (0)}  
\frac{d f_0^{K \pi} (t)}{d t} \bigg|_{t = 0}
\eeq
becomes smaller than the value (\ref{lambda0lin}) obtained in the linear 
approximation. Indeed, an analysis performed in the isospin limit and at 
first nonleading chiral order gave \cite{gl852} $\lambda_0^{K \pi} = 
0.017(4)$. 
In the meantime, the chiral perturbation series of $f_0^{K \pi} 
(t)$ has been pushed forward to the two-loop level 
\cite{PS02,BT03,Bijnens-Ghorbani07}.
Combined with an estimate of the relevant 
$\cO(p^6)$ low-energy couplings, the prediction \cite{CEEKPP05} 
$\lambda_0^{K \pi} = 0.013(3)$ was obtained. 

A further reduction
of the size of the slope parameter (compared to the next-to-leading 
order result) is also supported by approaches using dispersive methods. 
Based on a detailed study of strangeness-changing scalar form factors 
by Oller, Jamin and Pich \cite{JOP02}, they found \cite{JOP04} 
$\lambda_0^{K \pi} = 0.0157(10)$ and \cite{JOP06}          
$\lambda_0^{K \pi} = 0.0147(4)$, respectively. A similar result, 
$\lambda_0^{K \pi} = 0.01523(46) + 0.069 \Delta_{\rm CT}$, based on the 
dispersive representation (\ref{disprep}) was given in \cite{Passemar07}.
 
The present experimental situation is displayed in Table \ref{tab:1}.
Note that the numbers\footnote{The ISTRA+ result \cite{ISTRA} 
has been rescaled by $M_{\pi^+}^2 / M_{\pi^0}^2$.} shown here are those 
where a quadratic 
parametrization (\ref{quadpar}) has been used for the simultaneous 
determination of the vector form factor $f_+^{K \pi}(t)$.

\begin{table*}
\caption{Experimental results for $\lambda_0^{K \pi} \times 10^3$}
\label{tab:1}
\begin{tabular}{cccccc}
\hline\noalign{\smallskip}
ISTRA$+$ ($K^+_{\mu 3}$)& KTeV ($K_{L \mu 3}$) & KTeV ($K_{L \mu3} + K_{L e3}$) & NA48 ($K_{L \mu 3}$) & KLOE ($K_{L \mu 3}$) & KLOE ($K_{L \mu 3} + K_{L e3}$)\\
\noalign{\smallskip}\hline\noalign{\smallskip}
$\, 17.1 \pm 2.2 \,$ & $\, 12.8 \pm 1.8 \,$ & $\, 13.7 \pm 1.3 \,$ & $\, \,  9.5 \pm 1.4 \,$ 
& $\, \,  9.1 \pm 6.5 \,$ &
$\, 15.4 \pm 2.2 \,$ \\
\noalign{\smallskip}\hline
\end{tabular}
\end{table*}

It is difficult to accomodate the result of NA48 with those of the other
experiments\footnote{See also the critical review of the present data in 
\cite{FLAVIAnet}.}.
It is 
also hard to see \cite{Leutwyler07} how the outcome of NA48 could be 
reconciled with the results obtained in the dispersive analysis.
Furthermore, if the numbers given by ISTRA+ (obtained from $K^+$ 
decays) and NA48 (extracted from $K_{L}$ decays) were both true,  
this would signal an enormous isospin violation in the scalar form factors 
of $K_{\ell 3}$.

It is also an instructive exercise to estimate the scalar mass $M_S$ of 
the pole parametrization (\ref{polefit0}) by imposing the 
Callan-Treiman relation (\ref{Callan-Treiman}). Inserting
$f_0^{K \pi}(\Delta_{K \pi}) \simeq F_K / F_\pi$ 
in (\ref{polefit0}), one finds
\beq
M_S^2 \simeq \frac{\Delta_{K \pi}}{1 - 
F_\pi f_+^{K \pi}(0) / F_K} 
\simeq (1082 \,{\rm MeV})^2 ,
\eeq
being in reasonable agreement with the result found by KTeV 
\cite{KTeV}, $M_S = 1167(42) \, {\rm MeV}/c^2$, but rather 
far away from the value \cite{NA48}
$M_S = 1400(70) \, {\rm MeV}/c^2$ given by NA48.

\section{$\mbox{\boldmath $K_{\ell 3}$}$ form factors}
\label{sec: form factors}
\renewcommand{\theequation}{\arabic{section}.\arabic{equation}}
\setcounter{equation}{0}

In the following, we are using the notation introduced in \cite{CKNRT02}.
In particular, 
\beq
\wt{f}_{\pm}^{K^+ \pi^0} \! ,  \, \, \wt{f}_{\pm}^{K^0 \pi^-}
\eeq 
denote the pure QCD contributions (in principle at any order in the chiral 
expansion) plus the electromagnetic contributions to the meson masses and 
$\pi^0$--$\eta$ mixing. 
In fact 
\beq
{f}_{+}^{K^0 \pi^-} \! (0) \to
\wt{f}_{+}^{K^0 \pi^-} \! (0)
\eeq
has to be inserted in the master formula (\ref{master formula}) for the 
analysis of 
$K_{\ell 3}$ decays.  We also recall the
(lowest-order) expressions for the pseudoscalar masses
\beqa
 M^2_{\pi^\pm} &=& 2{B_0} \wh m + 2 e^2 Z {F_0}^2, \no \\
 M^2_{\pi^0} &=&  2{B_0} \wh m , \no \\
 M^2_{K^\pm} &=& {B_0}\left[ (m_s + \wh m) - \frac{2\ve^{(2)}}{\sqrt{3}}
(m_s - \wh m)\right] + 2e^2 Z {F_0}^2, \no \\
 M^2_{\stackrel{(-)}{K}{}^0} &=& {B_0} \left[(m_s + \wh m) + 
\frac{2\ve^{(2)}}{\sqrt{3}} (m_s - \wh m)\right] , \no \\
 M^2_\eta &=& \frac{4}{3} {B_0}\left( m_s + \frac{\wh m}{2}\right).
\label{treemass}
\eeqa
The mixing angle $\ve^{(2)}$ is given by
\beq
\label{epsilon}
\ve^{(2)} = \frac{\sqrt{3}}{4} \; \frac{m_d - m_u}{m_s - \wh m} ,
\eeq 
the symbol $\wh m$ stands for the mean value of the light quark masses,
\beq
\wh m = \frac{1}{2} (m_u + m_d),
\eeq
and ${B_0}$ is related to the vacuum condensate.
Finally, $M_{\pi}$ and $M_K$  denote the
isospin limits ($m_u = m_d$, $e = 0$) of the pion mass and the
kaon mass, respectively:
\beq
 M^2_{\pi} =  2{B_0} \wh m , \quad 
 M^2_K = {B_0} (m_s + \wh m).
\eeq

Whenever isospin-breaking effects are taken into account, the 
distinction of quantities like
\beqa
&&
\wt{f}_{+}^{K^+ \pi^0} \! \ne \wt{f}_{+}^{K^0 \pi^-} \! , \quad
\wt{f}_{-}^{K^+ \pi^0} \! \ne \wt{f}_{-}^{K^0 \pi^-} \! ,
\no \\
&&
M_\pi^\pm \ne M_{\pi^0} = M_\pi, \quad
M_{K^\pm} \ne M_{K^0} \ne M_K , \no
\eeqa
etc. has to be observed with meticulous care. 

To order $p^4$, $(m_d-m_u) p^2$, $e^2 p^2$, the $f_+$ form factors
are given by \cite{gl852,CKNRT02}
\begin{eqnarray}
 \wt{f}_{+}^{K^+\pi^0} \! (t) & = & 1 + \sqrt{3}   \Big( \ve^{(2)} + 
\ve^{(4)}_{\rm S} + \ve^{(4)}_{\rm EM} \Big) \nonumber \\*
&& {} +  \frac{1}{2} H_{K^+ \pi^0} (t) +  
 \frac{3}{2} H_{K^+ \eta} (t) +  H_{K^0 \pi^-} (t)    \nonumber \\*
&& {} +  \sqrt{3}  \ve^{(2)} \bigg[
 \frac{5}{2} H_{K \pi} (t) + \frac{1}{2} H_{K \eta} (t) \bigg]  ,
\label{ff1}
\end{eqnarray}
and
\begin{eqnarray}
\label{ff2}
 \wt{f}_{+}^{K^0 \pi^-} \! (t) & = & 1 + \frac{1}{2} H_{K^+ \pi^0} (t) +  
 \frac{3}{2} H_{K^+ \eta} (t) +  H_{K^0 \pi^-} (t)    \nonumber \\*
&&{} + \sqrt{3}  \ve^{(2)} \big[ H_{K \pi} (t) -  
H_{K \eta} (t) \big]  .
\end{eqnarray}
The loop functions $H_{PQ}(t)$ can be found in \cite{gl852,GL85}.
The quantity $\ve^{(4)}_{\rm S}$ is the strong contribution to the 
$\pi^0$--$\eta$ mixing angle arising at first nonleading order 
\cite{gl852} and 
$\ve^{(4)}_{\rm EM}$ is the corresponding term generated at $\cO(e^2 p^2)$ 
\cite{nr95}. The explicit expressions for $\ve^{(4)}_{\rm S}$ and 
$\ve^{(4)}_{\rm EM}$ can be found in \cite{CKNRT02}. 
We just note in passing that (\ref{ff1}) and (\ref{ff2}) 
imply the relation
\beq
\label{relp0}
\wt{f}_{+}^{K^+\pi^0} \! (0) =  
\wt{f}_{+}^{K^0 \pi^-} \! (0) 
\bigg[ 1 + \sqrt{3}   \Big( \ve^{(2)} + 
\ve^{(4)}_{\rm S} + \ve^{(4)}_{\rm EM} \Big) 
\bigg] ,
\eeq
which defines
\beq
\label{deltaSU2}
\delta_{\rm SU(2)} = \left\{
\begin{array}{ll} 0 & \, \mbox{for} \ K^0_{\ell 3} \\
2 \sqrt{3} 
\Big( \ve^{(2)} + \ve^{(4)}_{\rm S} + \ve^{(4)}_{\rm EM} \Big) 
& \, \mbox{for} \
K^+_{\ell 3} \end{array}
\right. 
\eeq
to the order $(m_d-m_u)p^2$, $e^2 p^2$.

The analogous expressions for the $f_{-}$ form factors are given by 
\cite{CKNRT02} 
\beqa
\wt{f}_{-}^{K^+ \pi^0} \! (t) &=&
\frac{4 \Delta_{K \pi}}{F_0^2} 
\bigg( 1 + \frac{\varepsilon^{(2)}}{\sqrt{3}} \bigg)  
\bigg[ L_5^r(\mu) - \frac{3}{256\pi^2} \ln\frac{M_{K^\pm}^2}{\mu^2} 
\bigg] 
\nonumber\\
&-& \frac{1}{128\pi^2F_0^2} \bigg[
\big( 3 + \sqrt{3} \varepsilon^{(2)} \big)
M_{\eta}^2\ln\frac{M_{\eta}^2}{M_{K^\pm}^2}
\nonumber\\
&&{}
\hphantom{\frac{1}{128\pi^2F_0^2} \bigg[ }
+ 2 \big( 3 - \sqrt{3} \varepsilon^{(2)}\big) 
M_{K^0}^2\ln\frac{M_{K^0}^2}{M_{K^\pm}^2} \nonumber\\
&&{}
\hphantom{\frac{1}{128\pi^2F_0^2} \bigg[ }
- 2 \big( 3 - \sqrt{3} \varepsilon^{(2)} \big) M_{\pi^\pm}^2\ln
\frac{M_{\pi^\pm}^2}{M_{K^\pm}^2}\,
\nonumber\\
&&{}
\hphantom{\frac{1}{128\pi^2F_0^2} \bigg[ }
+ \big( 1+3\sqrt{3} \varepsilon^{(2)} \big) M_{\pi^0}^2\ln
\frac{M_{\pi^0}^2}{M_{K^\pm}^2} \bigg]
\nonumber\\
&+&
\,\sum_{PQ}
\bigg\{ 
\bigg[ a_{PQ}(t) + \frac{\Delta_{PQ}}{2t} b_{PQ} \bigg] K_{PQ}(t) 
\nonumber\\
&&{} 
\hphantom{\sum_{PQ} \bigg\{  }
+ b_{PQ} \frac{F_0^2}{t} H_{PQ}(t) \bigg\}  . 
\label{fminustildep}
\eeqa
and 
\beqa
\wt{f}_{-}^{K^0 \pi^-} \! (t) &=&
\frac{4 \Delta_{K \pi}}{F_0^2}
\bigg( 1 + \frac{2  \varepsilon^{(2)}}{\sqrt{3}} \bigg)
\bigg[ 
L_5^r(\mu) - \frac{3}{256\pi^2} \ln \frac{M_{\pi^\pm}^2}{\mu^2} \bigg]
\nonumber\\
&-&
\frac{1}{128\pi^2F_0^2} \bigg[
2M_{K^0}^2\ln\frac{M_{K^0}^2}{M_{\pi^\pm}^2} 
\no \\
&&{}
\hphantom{ \frac{1}{128\pi^2F_0^2} \bigg[ }
+ \big( 3+2\sqrt{3} \varepsilon^{(2)} \big)
M_{\eta}^2\ln\frac{M_{\eta}^2}{M_{\pi^\pm}^2}
\no \\
&&{}
\hphantom{ \frac{1}{128\pi^2F_0^2} \bigg[ }
- \big( 3+2\sqrt{3} \varepsilon^{(2)} \big)
M_{\pi^0}^2\ln\frac{M_{\pi^0}^2}{M_{\pi^\pm}^2}
\bigg]
\nonumber\\
&+& 
\sum_{PQ}
\Bigg\{ 
\bigg[ c_{PQ}(t) + \frac{\Delta_{PQ}}{2t} d_{PQ} \bigg] K_{PQ}(t) 
\nonumber\\
&&{}
\hphantom{\sum_{PQ} \Bigg\{  }
+ d_{PQ}\,\frac{F_0^2}{t} H_{PQ}(t) \Bigg\}  ,
\label{fminustildez}
\eeqa
The sum runs over the meson pairs $K^+ \pi^0$, $K^0 \pi^+$, $K^+ \eta$ 
occurring in loop diagrams.  The function $K_{PQ} (t)$ is given by
\beq
K_{PQ} (t)  =   \frac{\Delta_{PQ}}{2 t} \bar{J}_{PQ} (t) ,
\eeq
with the short-hand notation 
$
\Delta_{P Q} = M_P^2 - M_Q^2
$
and the loop function $\bar{J}_{PQ} (t)$ defined in \cite{GL85}.
The coefficients
$a_{PQ} (t)$, $b_{PQ}$ 
$c_{PQ} (t)$, $d_{PQ}$ 
are listed in the Appendix.

In the next step we are trading the low-energy constant $L_5^r(\mu)$ 
appearing in the expressions for the $f_-$ form factors for the ratio
$F_K / F_\pi$. To this end we employ the relation \cite{lept}
\beqa
\frac{F_K}{F_\pi} &=& 
1 + \frac{4 \Delta_{K \pi}}{F_0^2} L_5^r(\mu) 
\bigg(1 - \frac{2 \ve^{(2)}}{\sqrt{3}} \bigg) \no \\
&& \mbox{} - \frac{1}{8(4\pi)^2 {F_0}^2} \bigg[ 3 M^2_{\eta} \ln 
\frac{M^2_{\eta}}{\mu^2}  
+ 2 M^2_{K} \ln \frac{M^2_{K}}{\mu^2}  \no \\
&& \mbox{} 
\hphantom{- \frac{1}{8(4\pi)^2 {F_0}^2} \bigg[} 
- 5 M^2_{\pi} \ln \frac{M^2_{\pi}}{\mu^2} 
\bigg] \no \\
&& \mbox{}
+\frac{\sqrt{3} \varepsilon^{(2)}}{4(4\pi)^2F_0^2}\Bigg[ 
M^2_{\eta} \ln \frac{M^2_{\eta}}{\mu^2} - 
M^2_{\pi} \ln \frac{M^2_{\pi}}{\mu^2} \no \\ 
&& \mbox{} 
\hphantom{+\frac{\sqrt{3} \varepsilon^{(2)}}{4(4\pi)^2F_0^2}\bigg[} 
+\frac{2}{3} \Delta_{K \pi} \bigg( \ln \frac{M^2_{K}}{\mu^2} + 
1 \bigg)
\Bigg] 
\label{FKdivFpi}\no \\
\eeqa
in (\ref{fminustildep}) and (\ref{fminustildez}).
In this way we find
\beqa
\label{fminKplus}
\wt{f}_{-}^{K^+ \pi^0} \! (t) &=&
\bigg( \frac{F_K}{F_\pi} - 1 \bigg)
\big( 1+\sqrt{3} \varepsilon^{(2)} \big)  
\nonumber\\
&-& \frac{\sqrt{3} \varepsilon^{(2)} }{(4 \pi F_0)^2}
\bigg( \Delta_{K \pi} - M_\pi^2 \ln \frac{M_K^2}{M_\pi^2} \bigg) \no \\
&+&  \frac{\Delta_{\pi^\pm \pi^0} }{4 (4 \pi F_0)^2}
\bigg(5 - 3 \ln \frac{M_K^2}{M_\pi^2} \bigg)   
\no \\
&+&
\sum_{PQ}
\Bigg\{ 
\bigg[ a_{PQ}(t) + \frac{\Delta_{PQ}}{2t} b_{PQ} \bigg] K_{PQ}(t) 
\nonumber\\
&&
\hphantom{\,\sum_{PQ}\bigg\{ \bigg[} 
+ b_{PQ}\,\frac{F_0^2}{t} H_{PQ}(t) \Bigg\} 
\eeqa
and
\beqa
\label{fminK0}
\wt{f}_{-}^{K^0 \pi^-} \! (t) &=&
\bigg( \frac{F_K}{F_\pi} - 1 \bigg) 
\bigg(1 + \frac{4 \ve^{(2)}}{\sqrt{3}} \bigg)
\no \\
&-& \frac{\ve^{(2)}}{\sqrt{3} (4 \pi F_0)^2} 
\Bigg( 
\Delta_{K \pi}
- M_\pi^2 \ln \frac{M_K^2}{M_\pi^2}
\Bigg) 
\no \\
&+& \frac{\Delta_{\pi^\pm \pi^0}}{4 (4 \pi F_0)^2} \no \\
&+& 
\sum_{PQ}
\Bigg\{ 
\bigg[ c_{PQ}(t) + \frac{\Delta_{PQ}}{2t} d_{PQ} \bigg] K_{PQ}(t) 
\nonumber\\
&&{}
\hphantom{\sum_{PQ} \Bigg\{  }
+ d_{PQ}\,\frac{F_0^2}{t} H_{PQ}(t) \Bigg\}  .
\eeqa

\section{Scalar form factors}
\label{sec: Scalar form factors}
\renewcommand{\theequation}{\arabic{section}.\arabic{equation}}
\setcounter{equation}{0}

In the combination
\beq
\label{f0p}
\wt{f}_0^{K^+ \pi^0} \! (t) = \wt{f}_+^{K^+ \pi^0} \! (t) + 
\frac{t}{\Delta_{K^+ \pi^0}} \wt{f}_-^{K^+ \pi^0} \! (t)
\eeq
the terms with $H_{P Q} (t)$ cancel completely because of the relation
\begin{eqnarray}
&& {}  \frac{1}{2} H_{K^+ \pi^0} (t) +  
 \frac{3}{2} H_{K^+ \eta} (t) +  H_{K^0 \pi^-} (t)    \nonumber \\*
&& {} +  \sqrt{3} \, \ve^{(2)} \bigg[
 \frac{5}{2} H_{K \pi} (t) + \frac{1}{2} H_{K \eta} (t) \bigg]  \no \\
&& {} + \sum\limits_{P Q} b_{P Q} F_0^2 H_{P Q} (t) /\Delta_{K^+ \pi^0} = 
0
\end{eqnarray}
This is to be expected
as $H_{P Q} (t)$ contains the low-energy coupling $L_9^r(\mu)$ being 
related to pure vector exchange.

Quite analogously, the terms with $H_{P Q} (t)$ disappear also in the  
combination
\beq
\label{f00}
\wt{f}_0^{K^0 \pi^-} \! (t) = \wt{f}_+^{K^0 \pi^-} \! (t) + 
\frac{t}{\Delta_{K^0 \pi^-}} \wt{f}_-^{K^0 \pi^-} \! (t),
\eeq
as a consequence of
\begin{eqnarray}
&& \frac{1}{2} H_{K^+ \pi^0} (t) +  
 \frac{3}{2} H_{K^+ \eta} (t) +  H_{K^0 \pi^-} (t)    \nonumber \\*
&&{} + \sqrt{3} \, \ve^{(2)} \big[ H_{K \pi} (t) -  
H_{K \eta} (t) \big]  \no \\
&&{} + \sum\limits_{P Q} d_{P Q} F_0^2 H_{P Q} (t) / \Delta_{K^0 \pi^-} = 
0 .   
\end{eqnarray}

The scalar form factor of $K^+_{\ell 3}$ can now be written as
\beqa
\label{f0Kp}
&& \wt{f}_0^{K^+ \pi^0} \! (t) = \wt{f}_0^{K^+ \pi^0} \! (0) \no \\
&&{} + \frac{t}{\Delta_{K^+ \pi^0}}
\Bigg\{ \bigg( \frac{F_K}{F_\pi} - 1 \bigg) \big(1 + \sqrt{3} 
\ve^{(2)}\big) 
\no \\
&&{} - \frac{\sqrt{3} \ve^{(2)}}{(4 \pi F_0)^2} 
\bigg( 
\Delta_{K \pi}
- M_\pi^2 \ln \frac{M_K^2}{M_\pi^2}
\bigg)
\no \\
&&{} 
+ \frac{\Delta_{\pi^\pm \pi^0}}{4 (4 \pi F_0)^2} 
\bigg(5 - 3 \ln \frac{M_K^2}{M_\pi^2} \bigg)
\no \\
&&{} +\sum\limits_{P Q} \bigg[ \frac{1}{2} a_{P Q}(0) \Delta_{P Q}
\bar{J}_{P Q}^{\prime} (0)
+ \frac{1}{8} b_{P Q} \Delta_{P Q}^2
\bar{J}_{P Q}^{\prime \prime} (0) \bigg] \Bigg\} \no \\
&&{} + \frac{1}{\Delta_{K^+ \pi^0}} 
\sum\limits_{P Q}
\Bigg\{\frac{1}{2} a_{P Q}^{\prime} (0) \Delta_{P Q} t 
\bar{J}_{P Q}(t) \no \\
&&{} + \frac{1}{2} a_{P Q}(0) \Delta_{P Q} \big[ 
\bar{J}_{P Q}(t)- t 
\bar{J}_{P Q}^{\prime} (0) \big] 
\no \\
&&{} + \frac{1}{4} b_{P Q} \Delta_{P Q}^2  
\frac{\bar{J}_{P Q}(t)- t 
\bar{J}_{P Q}^{\prime} (0) 
-t^2\bar{J}_{P Q}^{\prime \prime}(0) /2}
{t}
\Bigg\} .
\eeqa 
The last term in the curly brackets is of order $t^2$ and the 
first derivative of the form factor can be read off directly from the 
term $\sim t$.
The analogous expression for the $K^0_{\ell 3}$ scalar form factor is 
given by
\beqa
\label{f0K0}
&& \wt{f}_0^{K^0 \pi^-} \! (t) = \wt{f}_0^{K^0 \pi^-} \! (0) \no \\
&&{} + \frac{t}{\Delta_{K^0 \pi^-}}
\Bigg\{ \bigg( \frac{F_K}{F_\pi} - 1 \bigg) 
\bigg( 1 + \frac{4 \ve^{(2)}}{\sqrt{3}} \bigg)
\no \\
&&{} - \frac{\ve^{(2)}}{\sqrt{3} (4 \pi F_0)^2} 
\Bigg( 
\Delta_{K \pi}
- M_\pi^2 \ln \frac{M_K^2}{M_\pi^2}
\Bigg) 
+ \frac{\Delta_{\pi^\pm \pi^0}}{4 (4 \pi F_0)^2} \no \\
&&{} +\sum\limits_{P Q} \bigg[ \frac{1}{2} c_{P Q}(0) \Delta_{P Q}
\bar{J}_{P Q}^{\prime} (0)
+ \frac{1}{8} d_{P Q} \Delta_{P Q}^2
\bar{J}_{P Q}^{\prime \prime} (0) \bigg] \Bigg\} \no \\
&&{} + \frac{1}{\Delta_{K^0 \pi^-}} 
\sum\limits_{P Q}
\Bigg\{\frac{1}{2} c_{P Q}^{\prime} (0) \Delta_{P Q} t 
\bar{J}_{P Q}(t) \no \\
&&{} + \frac{1}{2} c_{P Q}(0) \Delta_{P Q} \big[ 
\bar{J}_{P Q}(t)- t 
\bar{J}_{P Q}^{\prime} (0) \big] 
\no \\
&&{} + \frac{1}{4} d_{P Q} \Delta_{P Q}^2  
\frac{\bar{J}_{P Q}(t)- t 
\bar{J}_{P Q}^{\prime} (0) 
-t^2\bar{J}_{P Q}^{\prime \prime}(0) /2}
{t}
\Bigg\} .
\eeqa 
In the further evaluation of the slope parameter 
\beq
\lambda_0^{K^+ \pi^0} := \frac{M_{\pi^+}^2}{\wt{f}_+^{K^+ \pi^0} \! (0)}  
\frac{d \wt{f}_0^{K^+ \pi^0} \! (t)}{d t} \bigg|_{t = 0} 
\eeq
from (\ref{f0Kp}) we use (\ref{relp0}) and arrive at
\beqa
\label{l0p}
&& \lambda_0^{K^+ \pi^0}  = 
\frac{M_{\pi^+}^2}{\Delta_{K^+ \pi^0}}
\no \\
&&{} \times
\Bigg\{
\frac{F_K}{F_\pi \wt{f}_+^{K^0 \pi^-} \! (0)} - \frac{1}{\wt{f}_+^{K^0 
\pi^-} \! (0)} 
\no \\
&&{} 
\hphantom{\times  }  
- \frac{\sqrt{3} \ve^{(2)}}{(4 \pi F_0)^2} 
\bigg( 
\Delta_{K \pi}
- M_\pi^2 \ln \frac{M_K^2}{M_\pi^2}
\bigg) \\
&&{} 
\hphantom{\times  }  
+ \frac{\Delta_{\pi^\pm \pi^0}}{4 (4 \pi F_0)^2} 
\bigg(5 - 3 \ln \frac{M_K^2}{M_\pi^2} \bigg)
+ \big(1 - \sqrt{3} \ve^{(2)} \big)
\no \\
&&{} 
\hphantom{\times  }  
\times 
\sum\limits_{P Q} \! \bigg[ \frac{1}{2} a_{P Q}(0) \Delta_{P Q}
\bar{J}_{P Q}^{\prime} (0) 
+ \frac{1}{8} b_{P Q} \Delta_{P Q}^2
\bar{J}_{P Q}^{\prime \prime} (0) \bigg]  
\Bigg\} .
\nonumber
\eeqa 
For
\beq
\lambda_0^{K^0 \pi^-} := \frac{M_{\pi^+}^2}{\wt{f}_+^{K^0 \pi^-} \! (0)}  
\frac{d \wt{f}_0^{K^0 \pi^-} \! (t)}{d t} \bigg|_{t = 0} 
\eeq
we obtain the result
\beqa
\label{l00}
&& \lambda_0^{K^0 \pi^-}  = 
\frac{M_{\pi^+}^2}{\Delta_{K^0 \pi^-}}
\no \\
&&{} \times
\Bigg\{
\bigg(
\frac{F_K}{F_\pi \wt{f}_+^{K^0 \pi^-} \! (0)} - \frac{1}{\wt{f}_+^{K^0 
\pi^-} \! (0)} 
\bigg)
\bigg(
1 + \frac{4 \ve^{(2)}}{\sqrt{3}} 
\bigg)
 \\
&&{} 
\hphantom{\times  }
- \frac{\ve^{(2)} / \sqrt{3}}{(4 \pi F_0)^2} 
\bigg[ 
\Delta_{K \pi}
- M_\pi^2 \ln \frac{M_K^2}{M_\pi^2}
\bigg] 
+ \frac{\Delta_{\pi^\pm \pi^0}}{4 (4 \pi F_0)^2} 
\no \\
&& {} 
\hphantom{\times  }
+\sum\limits_{P Q} \! \bigg[ \frac{1}{2} c_{P Q}(0) \Delta_{P Q}
\bar{J}_{P Q}^{\prime} (0)
+ \frac{1}{8} d_{P Q} \Delta_{P Q}^2
\bar{J}_{P Q}^{\prime \prime} (0) \bigg] 
\Bigg\} .
\nonumber
\eeqa 

\section{Callan-Treiman relations}
\label{sec: Callan-Treiman relations}
\renewcommand{\theequation}{\arabic{section}.\arabic{equation}}
\setcounter{equation}{0}

For the investigation of the Callan-Treiman relations in the presence of 
isospin breaking efffects, it is convenient to consider the ratios
\beq
\frac{\wt{f}_0^{K^+ \pi^0} \! (\Delta_{K^+ \pi^0} \! )}  
{\wt{f}_0^{K^+ \pi^0} \! (0)}
, \quad
\frac{\wt{f}_0^{K^0 \pi^-} \! (\Delta_{K^0 \pi^-} \! )}  
{\wt{f}_0^{K^0 \pi^-} \! (0)} .
\eeq
In the case of $K^+_{\ell 3}$ decays, we find
\beqa
&& \frac{\wt{f}_0^{K^+ \pi^0} \! (\Delta_{K^+ \pi^0} \! )} 
{\wt{f}_0^{K^+ \pi^0} \! (0)} 
= 
\frac{F_K}{F_\pi \wt{f}_+^{K^0 \pi^-} \! (0) }
\no \\ 
&&{} - \frac{\sqrt{3} \ve^{(2)} }{(4 \pi F_0)^2} 
\Bigg(
\Delta_{K \pi}
- M_\pi^2 \ln \frac{M_K^2}{M_\pi^2}
\Bigg) \\
&&{} + \frac{\Delta_{\pi^\pm \pi^0} }{4 (4 \pi F_0)^2} 
\bigg(5 - 3 \ln \frac{M_K^2}{M_\pi^2} \bigg)
+ \big( 1 + \sqrt{3} \ve^{(2)} \big)
\no \\
&&{}
\times
\sum\limits_{P Q} \! 
\bigg[ a_{P Q} (\Delta_{K^+ \pi^0}) 
+ \frac{\Delta_{P Q} b_{P Q}}{2 
\Delta_{K^+ \pi^0} } \bigg] \!
K_{P Q} (\Delta_{K^+ \pi^0} \! ) 
\nonumber
\eeqa 
A further evaluation of the coefficients $a_{P Q}(\Delta_{K^+ \pi^0})$, 
$b_{P Q}$ and  of $K_{P Q}(\Delta_{K^+ \pi^0})$ leads to the alternative 
form 
\beqa
\label{CTp}
&& \frac{\wt{f}_0^{K^+ \pi^0} \! (\Delta_{K^+ \pi^0} \! )} 
{\wt{f}_0^{K^+ \pi^0} \! (0)} 
= 
\frac{F_K}{F_\pi \wt{f}_+^{K^0 \pi^-} \! (0) }
\no \\
&&{}
- \frac{\sqrt{3} \ve^{(2)} }{(4 \pi F_0)^2} 
\Bigg(
\Delta_{K \pi}
- M_\pi^2 \ln \frac{M_K^2}{M_\pi^2}
\Bigg) \no \\
&&{} + \frac{\Delta_{\pi^\pm \pi^0} }{4 (4 \pi F_0)^2} 
\bigg(5 - 3 \ln \frac{M_K^2}{M_\pi^2} \bigg)
\\
&&{}
+ \frac{M_\pi^2}{2 F_0^2} 
\bigg( \! 1 + \frac{12 \ve^{(2)}}{\sqrt{3}}
- \frac{4 \ve^{(2)} M_K^2}{\sqrt{3}M_\pi^2}  \! 
\bigg)
\bar{J}_{K^+ \pi^0} (\Delta_{K^+ \pi^0}\!) 
\no \\
&&{}
- \frac{M_\pi^2}{F_0^2} \! 
\bigg( \! 1 \! + \! \frac{2 \ve^{(2)}}{\sqrt{3}} \!
+ \! \frac{4 \ve^{(2)}M_K^2}{\sqrt{3}M_\pi^2} \! 
- \! \frac{2 \Delta_{\pi^\pm \pi^0}}{\Delta_{K \pi}}
\! \bigg) \!
\bar{J}_{K^0 \pi^-} \! (\! \Delta_{K^+ \pi^0}\!) 
\no \\
&&{} 
- \frac{M_\pi^2}{6 F_0^2} \!
\bigg( \! 1 \! + \! \frac{8 \ve^{(2)}}{\sqrt{3}} \!
+ \! \frac{4 \ve^{(2)}M_K^2}{\sqrt{3}M_\pi^2} \! 
- \! \frac{4 \Delta_{\pi^\pm \pi^0}}{\! \Delta_{K \pi}} \! \bigg)
\! \bar{J}_{K^+ \eta} (\! \Delta_{K^+ \pi^0}\!) .
\nonumber 
\eeqa
We remark that the specific combination of terms
\beq
\label{goodcombp}
\delta_{\rm CT}^{K^+ \pi^0}
\! :=
\frac{\wt{f}_0^{K^+ \pi^0} \! (\Delta_{K^+ \pi^0} \! )} 
{\wt{f}_0^{K^+ \pi^0} \! (0)} 
- 
\frac{F_K}{F_\pi \wt{f}_+^{K^0 \pi^-} \! (0) }
\eeq
vanishes at lowest order\footnote{To lowest order, the second term is equal to one
and the form factor itself is independent of the momentum transfer.}. 
Contributions to this expression 
start only at NLO and constitute an appropriate measure for 
corrections to the Callan-Treiman relation 
in the presence of isospin violation. 
In contrast, the quantity
\beq
\label{badcomb}
\Delta_{\rm CT}^{K^+ \pi^0} \! = 
\wt{f}_0^{K^+ \pi^0} \! (\Delta_{K^+ \pi^0} \! ) 
-
\frac{F_K}{F_\pi} 
= \sqrt{3} \, \varepsilon^{(2)} + \ldots,
\eeq
receives a large (but trivial) contribution already at the tree 
level, making it less convenient for the discussion of deviations 
from the Callan-Treiman limit in the case $m_u \ne m_d$, $e \ne 0$.
Note also that the NLO contribution to (\ref{goodcombp}) can be expressed 
in terms of physical masses, the ratio $F_K / F_\pi \wt{f}_+^{K^0 \pi^-} 
\! (0)$, 
$\varepsilon^{(2)}$ and the pion decay constant, whereas the 
computation of (\ref{badcomb}) at NLO requires additional 
information on $\varepsilon_{\rm S}^{(4)}$ and $\varepsilon_{\rm 
EM}^{(4)}$ entering in (\ref{relp0}).

The analogous formulas in the case of $K^0_{\ell 3}$ are given by
\beqa
&& \frac{\wt{f}_0^{K^0 \pi^-} \! (\Delta_{K^0 \pi^-} \!)} 
{\wt{f}_0^{K^0 \pi^-} \! (0)} 
= \frac{F_K}{F_\pi \wt{f}_+^{K^0 \pi^-} \! (0) } 
\no \\
&&{}
+ \frac{4 \ve^{(2)}}{\sqrt{3}} 
\bigg( \frac{F_K}{F_\pi \wt{f}_+^{K^0 \pi^-} \! (0)} 
- \frac{1}{\wt{f}_+^{K^0 \pi^-} \! (0)} 
\bigg) 
\\
&&{} - \frac{\ve^{(2)}}{\sqrt{3} (4 \pi F_0)^2}
\bigg( \Delta_{K \pi}
- M_\pi^2 \ln \frac{M_K^2}{M_\pi^2}
\Bigg) 
+ \frac{\Delta_{\pi^\pm \pi^0}}{4 (4 \pi F_0)^2} \no \\
&&{} + \sum\limits_{P Q} \! \bigg[c_{P Q} (\Delta_{K^0 \pi^-}\!)
+ \frac{\Delta_{P Q} d_{P Q}}{2 \Delta_{K^0 \pi^-}} \bigg]
\! K_{P Q}(\Delta_{K^0 \pi^-}\!) ,
\nonumber
\eeqa 
and
\beqa
\label{CT0}
&& \frac{\wt{f}_0^{K^0 \pi^-} \! (\Delta_{K^0 \pi^-} \!)} 
{\wt{f}_0^{K^0 \pi^-} \! (0)} 
= \frac{F_K}{F_\pi \wt{f}_+^{K^0 \pi^-} \! (0) } 
\no \\
&&{}
+ \frac{4 \ve^{(2)}}{\sqrt{3}} 
\bigg( \frac{F_K}{F_\pi \wt{f}_+^{K^0 \pi^-} \! (0)} 
- \frac{1}{\wt{f}_+^{K^0 \pi^-} \! (0)} 
\bigg) 
\no \\
&&{} - \frac{\ve^{(2)}}{\sqrt{3} (4 \pi F_0)^2}
\bigg( \Delta_{K \pi}
- M_\pi^2 \ln \frac{M_K^2}{M_\pi^2}
\Bigg) 
+ \frac{\Delta_{\pi^\pm \pi^0}}{4 (4 \pi F_0)^2} \no \\
&&{} 
- \frac{M_\pi^2}{2 F_0^2} 
\bigg( 1 - \frac{2 \ve^{(2)}}{\sqrt{3}} 
+ \frac{2 \Delta_{\pi^\pm \pi^0} M_K^2}{\Delta_{K \pi} M_\pi^2} \bigg)
\bar{J}_{K^+ \pi^0} (\Delta_{K^0 \pi^-}\!) \no \\
&&{} - \frac{M_\pi^2}{6 F_0^2} 
\bigg( 1 + \frac{6 \ve^{(2)}}{\sqrt{3}} 
- \frac{2 \Delta_{\pi^\pm \pi^0}}{\Delta_{K \pi}} \bigg)
\bar{J}_{K^+ \eta} (\Delta_{K^0 \pi^-}\!) ,
\eeqa 
respectively. In the following, the quantity
\beq
\label{goodcomb0}
\delta_{\rm CT}^{K^0 \pi^-}
\! :=
\frac{\wt{f}_0^{K^0 \pi^-} \! (\Delta_{K^0 \pi^-} \!)} 
{\wt{f}_0^{K^0 \pi^-} \! (0)} 
- \frac{F_K}{F_\pi \wt{f}_+^{K^0 \pi^-} \! (0) } 
\eeq
will be used to measure the size of corrections to the Callan-Treiman 
relation in the case of neutral kaon decays.

In the isospin limit, (\ref{CTp}) as well as (\ref{CT0}) reduce to 
the well known result \cite{gl852}
\beq
f_0^{K \pi}  (\Delta_{K \pi}) = \frac{F_K}{F_\pi} 
- \frac{M_\pi^2}{6 F_0^2}
\bigg[ 
3 \bar{J}_{K \pi} (\Delta_{K \pi}) 
+ 
\bar{J}_{K \eta} (\Delta_{K \pi})
\bigg] .
\eeq

\section{Size of isospin breaking}
\label{sec: Size of isospin breaking}
\renewcommand{\theequation}{\arabic{section}.\arabic{equation}}
\setcounter{equation}{0}

The size of strong isospin violation is determined by the mixing angle 
$\ve^{(2)}$ defined in (\ref{epsilon}) or, equivalently, by the ratio of 
quark mass differences
\beq
\label{R}
R := \frac{m_s - \wh{m}}{m_d - m_u} .
\eeq
Up to corrections of order $m_q^2$, the double ratio
\beq
\label{Q2}
Q^2 := 
\frac{m_s^2 - \wh{m}^2}{m_d^2 - m_u^2} = R \frac{m_s / \wh{m} +1}{2} 
\eeq
is given by 
meson masses and a purely electromagnetic contribution \cite{GL85}: 
\beq
Q^2 = \frac{\Delta_{K \pi} M_K^2 \big(1 + \cO(m_q^2)\big)}{M_\pi^2 
\big[ \Delta_{K^0 K^+} + \Delta_{\pi^+ \pi^0}
- ( \Delta_{K^0 K^+} + \Delta_{\pi^+ \pi^0})_{\rm EM} \big]} .
\eeq
As a consequence of Dashen's theorem \cite{Dashen}, the electromagnetic 
term vanishes to lowest order $e^2 p^0$. It can be expressed through 
chiral logarithms and a certain combination of electromagnetic couplings
\cite{Urech,nr95}:
\beqa
&& (\Delta_{K^0 K^+} + \Delta_{\pi^+ \pi^0})_{\rm EM} = \no \\
&& {}
e^2 M_K^2 \Bigg[ \frac{1}{4 \pi^2} \bigg(3 \ln \frac{M_K^2}{\mu^2}
- 4 + 2 \ln \frac{M_K^2}{\mu^2} \bigg) \no \\
&& {} + \frac{4}{3} (K_5 + K_6)^r (\mu) - 8 (K_{10} + K_{11})^r (\mu) \no 
\\
&& {} + 16 Z L_5^r(\mu) \Bigg] + \cO(e^2 M_\pi^2) .
\eeqa
The numerical values of the electromagnetic coupling constants appearing 
in this expression have been determined by several authors 
\cite{Bijnens-Prades97,Moussallam97,Anant-Moussallam04}. 
Here we are using the most recent result by Ananthanarayan and Moussallam
\cite{Anant-Moussallam04}. They obtain a rather large 
deviation from Dashen's limit,
\beq   
(\Delta_{K^0 K^+} + \Delta_{\pi^+ \pi^0})_{\rm EM} =
-1.5 \, \Delta_{\pi^+ \pi^0}
\eeq
which corresponds to
\beq
\label{Qnum}
Q = 20.7 \pm 1.2 ,
\eeq
where we have added a rather generous error to account for higher order 
corrections.

For the determination of 
\beq
\label{Rexpr}
R = \frac{2 Q^2}{m_s / \wh{m} + 1}
\eeq
we also need information about the quark mass ratio $m_s / \wh{m}$
as our second input parameter. Employing different methods 
\cite{Leutwyler96},
typical values around $m_s/ \wh{m} \sim 24$ have been obtained in the 
literature. We want to corroborate this size of the quark mass ratio by a 
numerical update of the determination of $m_s / \wh{m}$ with a method 
proposed by Leutwyler \cite{Leutwyler97} using the decay widths of $\eta 
\to \gamma \gamma$ and $\eta^\prime \to \gamma \gamma$. Defining the 
parameters $c_{\eta}$ and $c_{\eta^\prime}$ by \cite{Leutwyler97}
\beq
\Gamma(P \to \gamma \gamma) = \frac{\alpha^2 M_P^3}{64 \pi^3 F_\pi^2} 
c_P^2 , 
\eeq
the experimental values for the decay widths given in \cite{PDG06}
correspond to $c_\eta = 0.992 \pm 0.025$ and $c_{\eta^\prime} = 1.245 \pm 
0.022$. The quark mass ratio can be obtained from the system of equations 
\cite{Leutwyler97} (see also \cite{Kaiser-Leutwyler98,Kaiser-Leutwyler00})
\beqa
F_\eta^8 c_\eta + F_{\eta^\prime}^8 c_{\eta^\prime} &=& F_\pi / \sqrt{3} 
\label{syst1} \\
(F_\eta^8)^2 + (F_{\eta^\prime}^8)^2 &=& \frac{4 F_K^2 - F_\pi^2}{3} 
\label{syst2} \\
(F_\eta^8)^2 M_\eta^2 + (F_{\eta^\prime}^8)^2 M_{\eta^\prime}^2 &=& 
\frac{8 F_K^2 M_K^2 m_s/\wh{m}}{3 (m_s/\wh{m} + 1)}  \\
&-& \frac{F_\pi^2 M_\pi^2 (2 m_s/\wh{m}-1)}{3} . \no \label{syst3}
\eeqa
Eq.(\ref{syst2}) can be written in the form 
\cite{Leutwyler97}
\beq
F_\eta^8 = F_8 \cos \vartheta_8, \quad F_{\eta^\prime}^8 = F_8 \sin 
\vartheta_8
\eeq
with
\beq
(F_8)^2 = \frac{4 F_K^2 - F_\pi^2}{3} .
\eeq
Using (\ref{syst1}), the observed values of $c_\eta$ and $c_{\eta^\prime}$ 
require $\vartheta_8 = - 22.0^\circ$. Inserting this in (\ref{syst3}) 
yields 
the quark mass ratio
\beq
\label{msdmh}
m_s / \wh{m} = 24.7 \pm 1.0 \pm 0.3 \pm 0.1 = 24.7 \pm 1.1,
\eeq
where the errors refer to the uncertainties of 
$\Gamma(\eta \to \gamma \gamma)$,  
$\Gamma(\eta^\prime \to \gamma \gamma)$ and $F_K / F_\pi$. This value is 
perfectly consistent with $m_s / \wh{m} = 24.4 \pm 1.5$ obtained in 
\cite{Leutwyler96} based on different arguments.

Combining (\ref{Qnum}) and (\ref{msdmh}), the relation (\ref{Rexpr}) 
finally gives
\beq
\label{Rnum}
R = 33.5 \pm 4.0 \pm 1.5 = 33.5 \pm 4.3 .
\eeq
A value for $R$ of this size has been suggested in \cite{ABT01}. Note 
however that a recent analysis of $\eta \to 3 \pi$ 
at 
the two-loop level \cite{eta3pi} favours the values $R = 42.2$ and $Q = 
23.2$.

The result in (\ref{Rnum}) corresponds to 
\beq
\label{epsnum}
\ve^{(2)} = (1.29 \pm 0.17) \times 10^{-2}
\eeq
and will be used in our subsequent numerical analysis.  
We also note that (\ref{epsnum}) leads to the numerical value
\beq
\delta_{\rm SU(2)} = 0.058(8)
\eeq
for the parameter (\ref{deltaSU2}) in $K_{\ell 3}$ decays.

\section{Numerics at $\mbox{\boldmath $\cO \big( p^4, (m_d-m_u) p^2, e^2 
p^2 \big)$ }$}
\label{sec: Numerics}
\renewcommand{\theequation}{\arabic{section}.\arabic{equation}}
\setcounter{equation}{0}

For $M_{\pi^\pm}$, $M_{\pi^0}$, $M_{K^\pm}$ and $M_{K^0}$ we take the 
PDG06 values \cite{PDG06}, but for $M_\eta$ we employ the Gell-Mann--Okubo  
relation \cite{GMO}
\beq
\label{GMO}
3 \Delta_{\eta K} = \Delta_{K \pi}.
\eeq
As this relation has to be used to arrive at the theoretical formulas of 
Sects. \ref{sec: Scalar form factors} and \ref{sec: Callan-Treiman 
relations}, this is the only unambiguous choice for $M_\eta$ at the 
considered chiral order.

With this set of masses, the numerical value for the ratio 
$F_K / F_\pi \wt{f}_+^{K^0 \pi^-} \! (0)$ given in (\ref{valmagrat}), the 
size of strong isospin breaking (\ref{epsnum})  and 
$F_0 \simeq F_\pi = 92.2 \, \rm MeV$ \cite{Moussallam05} as our input 
parameters, we obtain the following results for the slope parameters at 
the chiral order $p^4, (m_d - m_u) p^2, e^2 p^2$:
\beqa
\label{lambdanum}
\lambda_0^{K^0 \pi^-} &=& 
( \! \underbrace{16.64}_{m_u = m_d} 
+ \underbrace{0.17}_{m_u \ne m_d} + \underbrace{0.14}_{\rm EM} \, ) \times 
10^{-3} \nn
 &=& 16.95(40)(5) \times 10^{-3}, \\ 
\lambda_0^{K^+ \pi^0} &=&
( \! \underbrace{16.64}_{m_u = m_d} 
- \underbrace{0.12}_{m_u \ne m_d} - \underbrace{0.08}_{\rm EM} \, ) \times 
10^{-3} \nn
 &=& 16.44(39)(4) \times 10^{-3}.
\eeqa
The value in the limit of isospin conservation ($m_u = m_d$, $e=0$) and 
the contributions 
generated by strong isospin breaking and electromagnetism are displayed 
separately. The latter two pieces turn out to be of the same 
size. In the total results,
the first error refers to (\ref{valmagrat}) and the second one 
from error to (\ref{epsnum}). 
We see that both sources of isospin violation generate only 
tiny shifts with respect to the result in the in isospin limit, 
with a splitting of the two slope parameters given by
\beq
\label{difftot4} 
\Delta \lambda_0 := \lambda_0^{K^0 \pi^-} \! \! - \lambda_0^{K^+ \pi^0} = 
(5.1 \pm 0.9) \times 
10^{-4}.
\eeq

At the Callan-Treiman point we find
\beq
\label{ctnull}
\delta_{\rm CT}^{K^0 \pi^-} 
= 1.7(1)(7) \times 
10^{-3}
\eeq
for the quantity defined in (\ref{goodcomb0}).
The analogous measure (\ref{goodcombp}) of the corrections to the 
Callan-Treiman relation for the charged kaon decay mode 
is given by 
\beq
\label{ctplus}
\delta_{\rm CT}^{K^+ \pi^0} 
= -10.4(0)(7) \times 
10^{-3} .
\eeq
In both cases, the
first error is related to (\ref{valmagrat}), and the second one to  
(\ref{epsnum}). The corresponding value in the isospin limit is given 
by
\beq
\frac{f_0^{K \pi}  (\Delta_{K \pi}  )} 
{f_0^{K \pi}  (0)} 
- \frac{F_K}{F_\pi f_+^{K \pi}  (0) } = - 3.6 \times 10^{-3} .
\eeq
Switching off the electromagnetic contribution in (\ref{ctnull}) and 
(\ref{ctplus}) we obtain
\beq
\delta_{\rm CT}^{K^0 \pi^-} \Big|_{e = 0}
= 1.9 \times 10^{-3} ,
\quad
\delta_{\rm CT}^{K^+ \pi^0} \Big|_{e = 0}
= -9.9 
\times 
10^{-3} .
\eeq

\section{Analysis at $\mbox{\boldmath $\cO (p^6)$ }$}
\label{sec: p6}
\renewcommand{\theequation}{\arabic{section}.\arabic{equation}}
\setcounter{equation}{0}

In the isospin limit ($m_u = m_d$, $e=0$), the NNLO result for scalar form 
factors was given in the form \cite{BT03}
\beqa
\label{combhans}
f_0^{K \pi}(t) &+& \frac{t}{\Delta_{K \pi}} \bigg(1 - \frac{F_K}{F_\pi} 
\bigg) = 1 +\bar{\Delta}(t) + \Delta(0)
\nn
&-& \frac{8 \Delta_{K \pi}^2}{F_\pi^4} 
\big[ C_{12}^r(M_\rho)  + C_{34}^r (M_\rho) \big] 
\nn
&+&
\frac{8t \Delta_{K \pi}}{F_\pi^4} 
\big[ 2 C_{12}^r (M_\rho)  +C_{34}^r (M_\rho) \big]
\nn
&+&
\frac{16 t M_{\pi}^2}{F_\pi^4} 
\big[ 2 C_{12}^r (M_\rho)  + C_{34}^r (M_\rho) \big]
\nn
&-&
\frac{8 t^2}{F_\pi^4} 
C_{12}^r (M_\rho) 
.
\eeqa
Following the strategy proposed in \cite{CEEKPP05}, we pull out the 
tree-level pieces $\sim L_i^r \times L_j^r$ from $\bar{\Delta}(t)$ and 
$\Delta(0)$ by
defining\footnote{Note that terms $\sim L_4^r \times L_5^r, L_5^r \times 
L_6^r, L_5^r \times L_8^r$, etc.  cancel 
in the combination of terms entering in (\ref{combhans}).}
\beqa
D(0) &=& \Delta(0) - \frac{8 \Delta_{K \pi}^2}{F_\pi^4} L_5^r(M_\rho)^2, 
\label{D0}
\\
\bar{D}(t) &=& \bar{\Delta}(t) + \frac{8 t \Delta_{K \pi}}{F_\pi^4} 
L_5^r(M_\rho)^2.
\label{Dquer} 
\eeqa
The scalar form factor can now be written as
\beqa
\label{scalfp6}
f_0^{K \pi}(t) &=& 
f_+^{K \pi}(0) +  
\frac{t}{\Delta_{K \pi}} \bigg(\frac{F_K}{F_\pi} - 1 \bigg) 
\nn
&&{}
+\frac{8t \Delta_{K \pi}}{F_\pi^4} 
\big[ 2 C_{12}^r(M_\rho)  +C_{34}^r(M_\rho)  - L_5^r(M_\rho)^2 \big]
\nn
&&{}
+\frac{16 t M_{\pi}^2}{F_\pi^4} 
\big[ 2 C_{12}^r(M_\rho)  +C_{34}^r (M_\rho) \big]
\nn
&&{}
-\frac{8 t^2}{F_\pi^4} 
C_{12}^r(M_\rho)  +\bar{D}(t),
\eeqa
where
\beqa
f_+^{K \pi}(0) &=& 1+D(0)
\\
&-& 
\frac{8 \Delta_{K \pi}^2}{F_\pi^4} 
\big[ C_{12}^r(M_\rho)  + C_{34}^r(M_\rho)  - L_5^r(M_\rho)^2 \big] 
. \nonumber
\eeqa
The expression for the normalized scalar form factor takes the form
\beqa
\label{scalnormp6}
\frac{f_0^{K \pi}(t)}{f_+^{K \pi}(0)}
=
1 &+&  
\frac{t}{\Delta_{K \pi}} \bigg(\frac{F_K}{F_\pi f_+^{K \pi}(0)} - 
\frac{1}{1 + D(0)} \bigg) 
\nn
&+&
\frac{8t (\Delta_{K \pi} - t)}{F_\pi^4} 
C_{12}^r(M_\rho)  
\nn
&+&
\frac{16 t M_{\pi}^2}{F_\pi^4} 
\big[ 2 C_{12}^r(M_\rho)  +C_{34}^r (M_\rho) \big]
\nn
&+& \frac{\bar{D}(t)}{1 + D(0)},
\eeqa
allowing the following conclusion:
Apart from the very small contribution 
\beq
\label{DeltaCTp6}
\frac{16 t M_{\pi}^2}{F_\pi^4} 
\big[ 2 C_{12}^r (M_\rho)  +C_{34}^r (M_\rho) \big]
=
\Delta_{\rm CT}^{{\rm tree}, \, p^6} \frac{t}{\Delta_{K \pi}},
\eeq
which is suppressed by a factor $M_\pi^2 / M_K^2$, the 
slope as well as the curvature of (\ref{scalnormp6}) depend 
only on the counterterm $C_{12}^r(M_\rho)$ if
the loop functions $\bar{D}(t)$, $D(0)$ are known and the quantity 
$F_K / F_\pi f_+^{K \pi}(0)$ is used as input parameter. 

Taking $\Delta (0)$, $\bar{\Delta} (t)$ ($K^0_{\ell 3}$) from \cite{BT03} 
and $L_5^r(M_\rho)$ (fit 10) from \cite{ABT01}, (\ref{D0}) and 
(\ref{Dquer}) assume the numerical values
\beqa
\label{D0num}
D(0) &=& -0.0134 \pm 0.0005, \\
\label{Dbar}
\bar{D} (t) &=& -0.23407 \, t/{\rm GeV}^2 
+0.833045 \, (t/{\rm GeV}^2)^2 \nn
&&{}
+1.25252 \, (t/{\rm GeV}^2)^3 .
\eeqa  

The relevant $p^6$ counterterms have been determined by using the 
$1/N_C$ expansion and truncating the hadronic spectrum to the lowest 
lying resonances \cite{CEEKPP05}. In this framework, the leading term in 
the large-$N_C$ expansion of the relevant couplings 
can be expressed in terms of the scalar and pseudoscalar octet masses 
($M_S$ and $M_P$) and the pion decay constant \cite{CEEKPP05}: 
\beqa
\label{Cres}
L_5^{\cS \cP} &=& \frac{F_\pi^2}{4 M_S^2}, \quad
C_{12}^{\cS \cP} = - \frac{F_\pi^4}{8 M_S^4}, \nn 
C_{34}^{\cS \cP} &=&  \frac{3 F_\pi^4}{16 M_S^4}+ \frac{F_\pi^4}{16 M_S^4} 
\bigg(1 - \frac{M_S^2}{M_P^2} \bigg)^2.
\eeqa
One assumes that the expressions given in (\ref{Cres}) determine the 
corresponding renormalized coupling constants at some typical hadronic 
matching scale $\mu$:
\beq
C_i^r(\mu) = C_i^{\cS \cP} .
\eeq
The value of the coupling constant at our standard reference scale 
$M_\rho$ can then be obtained by
\beq
C_i^r(M_\rho) = C_i^r(\mu) + \delta C_i (\mu, M_\rho),
\eeq
where
\beqa
\label{deltaC}
\delta C_i (\mu, M_\rho) &=&
\frac{1}{(4 \pi)^2} 
\Bigg\{ 
\frac{\Gamma_i^{(2)}}{(4 \pi)^2} \bigg( \ln \frac{\mu}{M_\rho} \bigg)^2 
\nn
&&{}
-
\big[ 2 \Gamma_i^{(1)} + \Gamma_i^{(L)} (M_\rho) 
\big] \ln \frac{\mu}{M_\rho} 
\Bigg\} 
\eeqa
is determined by the renormalization group equations \cite{BCE00}
\beqa
\mu \frac{\partial C_i^r (\mu)}{\partial \mu} &=& \frac{1}{(4 \pi)^2} 
\big[ 2 \Gamma_i^{(1)} + \Gamma_i^{(L)} (\mu) \big], \nn
\mu \frac{\partial \Gamma_i^{(L)} (\mu)}{\partial \mu} 
&=& 
- \frac{\Gamma_i^{(2)}}{8 \pi^2} .
\eeqa
For our analysis we need \cite{BCE00}
\beqa
\Gamma_{12}^{(2)} &=& \frac{19}{64}, 
\quad \Gamma_{12}^{(1)} = - \frac{13}{768 (4 \pi)^2}, 
\nn
\label{Gamma12L}
\Gamma_{12}^{(L)} &=& \frac{2}{3} L_1^r + \frac{4}{3} L_2^r + \frac{8}{9} 
L_3^r + \frac{3}{4} L_5^r
\eeqa
and
\beqa
\Gamma_{34}^{(2)} &=& -\frac{13}{32}, 
\quad \Gamma_{34}^{(1)} = - \frac{31}{2304 (4 \pi)^2}, 
\nn
\label{Gamma34L}
\Gamma_{34}^{(L)} &=& -L_1^r - \frac{3}{2} L_2^r - \frac{11}{12} 
L_3^r + L_4^r -\frac{3}{2} L_5^r.
\eeqa

The analysis of \cite{CPEN03} (scenario A) suggests the value 
$M_S = 1.48 \, \rm GeV$ for the lightest scalar nonet that survives the 
large-$N_C$ limit. 
With this choice of the mass parameter one gets
\beq
L_5^{\cS \cP} = 0.97 \times 10^{-3},
\eeq
which agrees exactly with the mean-value of $L_5^r (M_\rho)$ 
obtained in fit 10 of \cite{ABT01}.
For the pseudoscalar mass parameter, spectroscopy and chiral 
symmetry \cite{PDG06,CPEN03} suggest the value $M_P = 1.3 \, \rm GeV$. 
With this input one obtains\footnote{These numbers can be compared with 
previous determinations in \cite{Bijnens-Dhonte03,JOP04,CEEKPP05}.} 
\beq
\label{C12}
C_{12}^r (M_\rho) = (- 1.9^{+ 2.0}_{-0.4}) \times 10^{-6}
\eeq
and
\beq
\label{C34}
C_{34}^r (M_\rho) = (2.9^{+1.3}_{-5.0}) \times 10^{-6}.
\eeq
The errors were estimated by calculating (\ref{deltaC}) using
(\ref{Gamma12L}) and 
(\ref{Gamma34L}), respectively.
The numerical values (together with their errors) of the 
$L_i^r (M_\rho)$ were taken from fit 10 of 
\cite{ABT01}. The matching scale $\mu$ was varied between 
$M_\eta$ 
and $\rm 1 \, GeV$ which
provides us with an estimate of the intrinsic uncertainty 
due to subleading contributions in $1/N_C$. 
Note that the asymmetric errors in (\ref{C12}) and (\ref{C34}) 
originate from the quadratic term
in (\ref{deltaC}) as a consequence of the two-loop renormalization
group equation.

Expanding the scalar form factor as
\beq
\label{quadparf0}
\frac{f_0^{K\pi} (t)}{f_+^{K\pi} (0)}
= 
1 + \lambda_0^{K \pi} \frac{t}{M_{\pi^+}^2}
+ \frac{1}{2} c_0^{K \pi } \big(\frac{t}{M_{\pi^+}^2} \Big)^2 
+ \ldots,
\eeq
(\ref{scalnormp6}) implies
\beqa
\label{l0p6}
\lambda_0^{K \pi} = M_{\pi^+}^2 &\bigg\{&
\frac{1}{\Delta_{K \pi}} \bigg(\frac{F_K}{F_\pi f_+^{K \pi}(0)} - 
\frac{1}{1 + D(0)} \bigg) 
\nn
&+&
\frac{8 \Delta_{K \pi} }{F_\pi^4} 
C_{12}^r(M_\rho)  
\nn
&+&
\frac{16 M_{\pi}^2}{F_\pi^4} 
\big[ 2 C_{12}^r(M_\rho)  +C_{34}^r (M_\rho) \big]
\nn
&+& \frac{\bar{D}^\prime(0)}{1 + D(0)} \bigg\}
\eeqa
for the slope parameter.
Using the two-loop results (\ref{D0num}),  (\ref{Dbar}) and  estimating 
the pertinent combination of 
low-energy couplings in the way described above, we find 
\beq
\label{lambda0res}
\lambda_0^{K \pi} = 
\big( 13.9^{+1.3}_{-0.4} \pm 0.4 \big) \times 10^{-3}.
\eeq
The first error is related to the uncertainties in the 
determination of the $C_i$ and the second one to those in  $F_K / F_\pi 
f_+(0)$ and $D(0)$. 

The expression for the curvature reads
\beq
\label{c0p6}
c_0^{K \pi} = M_{\pi^+}^4 \bigg\{
- \frac{16  }{F_\pi^4} C_{12}^r(M_\rho)  
+ \frac{\bar{D}^{\prime \prime}(0)}{1 + D(0)} 
\bigg\},
\eeq
which leads to the predicion
\beq
\label{curvres}
c_0^{K \pi } = \big( 8.0^{+0.3}_{-1.7} \big) \times 10^{-4},
\eeq
once (\ref{Dbar}) together with (\ref{C12}) have been inserted.
Note that the naive pole parametrization (\ref{polefit0}) would predict
\beq
c_0^{K \pi } \big|_{\rm pole \, \, fit} = 2 \big( \lambda_0^{K \pi} 
\big)^2 
\simeq 4 \times 10^{-4},
\eeq
where the  numerical value was obtained by inserting 
(\ref{lambda0res}).
This exercise shows that the pole fit is unable to reproduce the 
curvature found
in (\ref{curvres}). It was already noticed in \cite{BOPS06} that the 
pole parametrization underestimates the curvature derived from dispersion 
theory when realistic values of the slope and the pole mass are assumed 
(see also the discussion in the next section).

The combination of counterterms entering in (\ref{DeltaCTp6}) is given by
\beq
2 C_{12}^r (M_\rho) + C_{34}^r (M_\rho) = 
\big( -0.9^{+3.8}_{-3.4} \big) \times 10^{- 6}
\eeq
which translates into 
\beq
\label{DeltaCTtree}
\Delta_{\rm CT}^{{\rm tree}, \, p^6}  =   
\big( -0.8^{+3.5}_{-3.1} \big) \times 10^{-3}.
\eeq
Combined with the two-loop result given in \cite{Bijnens-Ghorbani07}, the 
total $p^6$ result (in the isospin-limit) reads
\beq
\Delta_{\rm CT}  =    
\big( -7.0^{+3.5}_{-3.1} \big) \times 10^{-3}.
\eeq 

For completeness, we also note 
\beq
C_{12}^r (M_\rho) + C_{34}^r (M_\rho) - L_5^r(M_\rho)^2 = 
\big( 0.1^{+1.1}_{-1.2}  \big) \times 10^{- 6},
\eeq
which determines the size of $f_+^{K \pi}(0)$
as 
\beq
\label{f0num}
f_+^{K \pi}(0) = 0.986 \pm 0.007_{1/N_c} \pm 0.002_{M_S, M_P} .
\eeq 
Apart from varying the matching scale, we have also added a second error 
to account for the uncertainty in the choice of the 
resonance masses, as our central value for $f_{p^6}^{\rm tree}$ given by
\beq
f_{p^6}^{\rm tree} = -\frac{\Delta_{K \pi}^2}{2 M_S^4} \bigg(1 - 
\frac{M_S^2}{M_P^2} \bigg)^2 
\eeq
depends strongly on the (relative) size of the mass parameters.
The number given in (\ref{f0num}) is 
to be compared with the still currently used Leutwyler-Roos value 
$f_+^{K \pi}(0) = 0.961 \pm 0.008$ \cite{Leutwyler-Roos}.

\section{Dispersive analysis}
\label{sec: Dispersive Analysis}
\renewcommand{\theequation}{\arabic{section}.\arabic{equation}}
\setcounter{equation}{0}

Employing the dispersive representation (\ref{disprep}) of the scalar form 
factor used in \cite{BOPS06,Passemar07,Bernard-Passemar07}, the slope 
parameter can be written as
\beq
\lambda_0^{K \pi} = \frac{M_{\pi^+}^2}{\Delta_{K \pi}} 
\big( \ln C - G(0) \big).
\eeq
Evaluating (\ref{disprep}) at $t= \Delta_{K \pi}$, one finds the relation 
\beq
C = \frac{F_K}{F_\pi f_+^{K \pi}(0)} +
\frac{\Delta_{\rm CT}}{f_+^{K \pi}(0)}.
\eeq 
Using (\ref{valmagrat}) and the estimate $\pm 0.01$ for the uncertainty 
due to $\Delta_{\rm CT} /f_+^{K \pi}(0)$, the parameter $C$ assumes the 
value
\beq
C = 1.2424 \pm 0.0045 \pm 0.01,
\eeq
or, equivalently,
\beq
\ln C = 0.2170 \pm 0.0036 \pm 0.0080.
\eeq
Together with \cite{BOPS06}
\beq
G(0) = 0.0398 \pm 0.0036 \pm 0.002,
\eeq
the dispersive analysis gives
\beq
\label{l0numdisp}
\lambda_0^{K \pi} = (15.1 \pm 0.8) \times 10^{-3}
\eeq
for the slope parameter,
being consistent with the corresponding result (\ref{lambda0res})   
based on resonance saturation. This number is also in good agreement with 
other results using a dispersive approach \cite{JOP06,JOP04}.

The expression for the curvature reads \cite{BOPS06}
\beqa
c_0^{K \pi } &=& (\lambda_0^{K \pi})^2
- \frac{2 M_{\pi^+}^4 G^\prime (0)}{\Delta_{K \pi}} \nn
 &=&
(\lambda_0^{K \pi})^2 +(4.16 \pm 0.50) \times 10^{-4}.
\eeqa
Inserting (\ref{l0numdisp}), the curvature is given by
\beq
\label{codispnum}
c_0^{K \pi } = (6.4 \pm 0.6) \times 10^{-4},
\eeq
which is again consistent with the result (\ref{curvres}) of the previous 
analysis. 

Translated in terms of the size of the low-energy constant 
$C_{12}^r (M_\rho)$, the dispersive analysis seems to favour values at the 
upper end of the range suggested by (\ref{C12}). Explicitly,
(\ref{codispnum}) corresponds to
\beq
C_{12}^r (M_\rho) = (0.0 \pm 0.7) \times 10^{-6}.
\eeq

An analogous comparison can be made using the recent results \cite{JOP06}
of the dispersive approach of Jamin, Oller and Pich \cite{JOP02}. Their method
is based on a coupled-channel solution 
of the dispersion relation for the scalar form factor
which includes also the $K \eta^\prime$ channel. They find \cite{JOP06}
\begin{eqnarray}
\frac{d}{dt}
\frac{f_0^{K \pi} (t)}{f_0^{K \pi}(0)}
\bigg|_{t=0}
&=&
0.773(21) \, {\rm GeV}^{-2}, \\
\frac{d^2}{dt^2}
\frac{f_0^{K \pi} (t)}{f_0^{K \pi}(0)}
\bigg|_{t=0}
&=&
1.599(52) \, {\rm GeV}^{-4},
\end{eqnarray}
which corresponds to the values 
\begin{equation}
\lambda_0^{K \pi} = (14.7 \pm 0.4) \times 10^{-3},
\end{equation}
and
\begin{equation}
c_0^{K \pi } = (6.07 \pm 0.20) \times 10^{-4}.
\end{equation}

\section{Contributions of order $\mbox{\boldmath $(m_d - m_u) 
p^4$ }$}
\label{sec: p6isobreak}
\renewcommand{\theequation}{\arabic{section}.\arabic{equation}}
\setcounter{equation}{0}

Recently, isospin breaking in the $K_{\ell 3}$ form factors has also been 
studied at the two-loop level \cite{Bijnens-Ghorbani07}. The results for 
the scalar form factor of $K^0 \to \pi^- \ell^+ \nu_{\ell}$ with $C_i^r = 
0$ turn out to be essentially the same those in the isospin limit. From 
Fig. 13 of \cite{Bijnens-Ghorbani07} one extracts 
\beq
\label{Schleifenbeitrag}
\Delta \lambda_0 \big|_{C_i^r = e = 
0}  \simeq 5 \times 10^{- 4} 
\eeq
The remaining contribution containing the $p^6$ low-energy 
couplings $C_i^r$ is given by
\beqa
\label{diffCi}
\lefteqn{
\Delta \lambda_0 \big|_{C_i^r}  =
\frac{32 \varepsilon^{(2)} \Delta_{K \pi} M_{\pi^+}^2}{\sqrt{3} F_\pi^4}
}
\nn
 & & \times 
\big( 2 C_{12} + 6 C_{17} + 6 C_{18} + 3 C_{34} + 3 C_{35} \big)^r 
(M_\rho).
\eeqa
This expression can be obtained by combining the pertinent terms given 
in Appendix B 
of \cite{Bijnens-Ghorbani07}. Resonance estimates of the low-energy 
couplings appearing in (\ref{diffCi}) were given in \cite{CEEKPP06}. Using 
these results, we find\footnote{Additional contributions due to 
$\eta^\prime$ exchange 
\cite{Kaiser07} cancel in the combination $2 C_{18} + C_{35}$.}
\beqa
\label{combresest}
\lefteqn{
\big( 2 C_{12} + 6 C_{17} + 6 C_{18} + 3 C_{34} + 3 C_{35} 
\big)^{\mathcal{S} \mathcal{P}} 
}
\nn
 & & =
\frac{F_\pi^4}{4 M_S^4} 
\bigg(1 - \frac{3 M_S^2}{2 M_P^2} - 
\frac{M_S^2}{M_{\eta^\prime}^2} 
+ 6 \lambda_2^{\mathcal{S} \mathcal{S}} \bigg).
\eeqa
Using our standard values for the resonance mases $M_S$, $M_P$ and our 
usual determination of the uncertainty of the large $N_C$ estimate, we 
find
\beqa
\lefteqn{
\big( 2 C_{12} + 6 C_{17} + 6 C_{18} + 3 C_{34} + 3 C_{35} 
\big)^r (M_\rho) 
}
\nn
 & & =
\big( -1.25 + 2.26 \lambda_2^{\mathcal{S} \mathcal{S}}
\pm 0.7_{1/N_C} \big) \times 10^{-5}. 
\eeqa
No estimates of the parameter $\lambda_2^{\mathcal{S} \mathcal{S}}$ are 
available so far, but in analogy to 
\cite{CEEKPP06} 
$\lambda_1^{\mathcal{S} \mathcal{P}} = - d_m / c_m$ 
one may expect \cite{Ecker-private}
\beq
\big| \lambda_2^{\mathcal{S} \mathcal{S}} \big| \lets 1.
\eeq
Varying $\lambda_2^{\mathcal{S} \mathcal{S}}$ in this interval, 
adding (\ref{Schleifenbeitrag}) and the 
electromagnetic contribution discussed in Sect. \ref{sec: Numerics}, we 
expect the total value for the difference 
of the two slope parameters to lie within the rather small range
\beq
0
\lets
\Delta \lambda_0 
\lets 
10^{-3} .
\eeq

The two-loop contributions to the correction terms of the 
Callan-Treiman relation in the presence of isospin violation were also 
given in \cite{Bijnens-Ghorbani07}. Translated in terms of the quantities 
defined in (\ref{goodcomb0}) and (\ref{goodcombp}), they find
\beq
\label{CTBG0}
\delta_{\rm CT}^{K^0 \pi^-}
\Big|_{C_i^r = e = 0}
=
-5.6 \times 10^{-3}
\eeq
and
\beq
\label{CTBGp}
\delta_{\rm CT}^{K^+ \pi^0}
\Big|_{C_i^r = e = 0}
= 
-13.3 \times 10^{-3},
\eeq
respectively.
These results should be supplemented by the associated local contributions 
arising at this order \cite{Bijnens-Ghorbani07}, which are, however, also 
plagued by partly 
undetermined low-energy couplings. We demonstrate this only for the  
purely isospin violating combination
\beqa
\lefteqn{
\label{CTdiff}
\Big(
\delta_{\rm CT}^{K^0 \pi^-}
\! \! -
\delta_{\rm CT}^{K^+ \pi^0} 
\Big) 
\Big|_{C_i^r}
=
\frac{32 \varepsilon^{(2)} M_K^4}{\sqrt{3} F_\pi^4}
\big( 2 C_{12} + 2 C_{14}
}
\nn
 & & {}
+ 2 C_{15} + 6 C_{17} + 6 C_{18} + 4 C_{34} + 3 
C_{35} \big)^r 
(M_\rho),
\eeqa
where terms $\sim \varepsilon^{(2)} M_\pi^2$ have been discarded.
In addition to the undetermined parameter 
$\lambda_2^{\mathcal{S} \mathcal{S}}$ already encountered in 
(\ref{combresest}), the resonance estimate for the $p^6$ 
low-energy coupling $C_{14}$ is still incomplete \cite{CEEKPP06},
preventing a reliable numerical determination of (\ref{CTdiff}) 
(and even more for the individual terms)
for the time being.

Nevertheless, based on the numbers (\ref{ctnull}) and (\ref{ctplus}) 
found at NLO, the partial NNLO
results shown in (\ref{CTBG0}) and (\ref{CTBGp}), our estimate 
of the isospin symmetric local $p^6$ contribution 
(\ref{DeltaCTtree}) 
and a rough 
order-of-magnitude estimate of 
not yet determined local terms of the order $(m_d-m_u) p^4$ (a typical 
term is shown in (\ref{CTdiff})), 
we expect numerically small corrections to the Callan-Treiman relation 
also in the 
presence of isospin violation with
\beq
\big| \delta_{\rm CT}^{K^0 \pi^-} \big|, \,
\big| \delta_{\rm CT}^{K^+ \pi^0} \big| \lets 10^{-2} .
\eeq

\section{Summary and conclusions}
\label{sec: Conclusions}
\renewcommand{\theequation}{\arabic{section}.\arabic{equation}}
\setcounter{equation}{0}

In this work we have discussed the theoretical predictions for the scalar 
form factors of $K_{\ell 3}$ decays within the standard model. 

The leading non-vanishing contribution to the scalar slope 
arises at order $p^4$ in the chiral expansion. The theoretical 
expression for the scalar form factor was worked 
out already more than twenty years ago \cite{gl852} in the limit of 
isospin conservation. In this case, the slope parameter is 
uniquely determined by the pseudoscalar masses, the pion decay constant 
and the ratio
\beq
F_K \big/ F_\pi \wt{f}_+^{K^0 \pi^-} \! (0) = 1.2424(45).
\eeq
The remarkably precise numerical value given here can be obtained by 
combining the 
latest experimental data on $K_{\mu 2(\gamma)}$, $\pi_{\mu 2 (\gamma)}$, 
$K_{Le3}^0$ and $V_{ud}$ with the corresponding theoretical expressions. 
Using this input, one finds 
\beq
\lambda_0^{K \pi} \big|_{p^4} = (16.64 \pm 0.39) \times 10^{-3}.
\eeq

The isospin violating contributions of order $(m_d - m_u) p^2$ and 
$e^2 p^2$ to the $K_{\ell 3}$ form factors were considered for the first 
time in \cite{CKNRT02}. The effects of strong isospin breaking are 
proportional to the mixing angle
\beq
\ve^{(2)} = \frac{\sqrt{3}}{4} \; \frac{m_d - m_u}{m_s - \wh m} 
= (1.29 \pm 0.17) \times 10^{-2}.
\eeq 
The numerical value shown here was obtained by using the corrections 
to Dashen's limit given in \cite{Anant-Moussallam04}. 
The electromagnetic contributions of order $e^2 p^2$ entering in the slope 
parameters $\lambda_0^{K^0 \pi^-}$ and $\lambda_0^{K^+ \pi^0}$ can be expressed 
through the electromagnetic pieces of the pseudoscalar masses as well as 
the coupling $Z$ associated the chiral Lagrangian of order $e^2 p^0$, 
which can also be related to the pion mass difference (to the considered 
order). Both sources of isospin violation generate only  a tiny shift  of 
the two slope parameters  (compared to the isospin symmetric limit) with 
a splitting 
$\Delta \lambda_0 = \lambda_0^{K^0 \pi^-} \! \! - \lambda_0^{K^+ \pi^0}$ 
given by
\beq
\Delta \lambda_0 \big|_{(m_d-m_u)p^2, e^2 p^2} 
= 
(5.1 \pm 0.9) \times 
10^{-4}
\eeq
at this chiral order.

The corrections arising at order $p^6$ (in the isospin limit) turn out to 
be quite sizeable. Combining 
the two-loop results of chiral perturbation theory \cite{BT03} with an 
updated estimate of the necessary $p^6$ low-energy couplings, the numerical
value of the slope parameter in the isospin symmetric limit is given by 
\beq
\label{chptres}
\lambda_0^{K \pi} = 
\big( 13.9^{+1.3}_{-0.4} \pm 0.4 \big) \times 10^{-3}.
\eeq
The main uncertainty in this result comes from a certain combination of 
$p^6$ low energy couplings which has been determined by an updated analysis 
based on \cite{CEEKPP05,CPEN03}.

Using the dispersive representation proposed in \cite{BOPS06} with 
(\ref{valmagrat}), we find 
\beq
\lambda_0^{K \pi} = (15.1 \pm 0.8) \times 10^{-3},
\eeq
being in good agreement with the value (\ref{chptres}) obtained in  
chiral perturbation theory and also with other results \cite{JOP06,JOP04} 
using dispersion techniques.

The inclusion of isospin violating contributions of order $(m_d - m_u) 
p^4$ does not change this picture substantially. We expect an additional 
uncertainty for the values of the slope parameters of at most $\pm 
10^{-3}$, mainly due to not yet fully determined low-energy couplings. 
Combining the two-loop results given in \cite{Bijnens-Ghorbani07} with an 
estimate of a further combination of low-energy couplings, the difference 
of the two slope parameters should be confined to the rather small range
\beq
0 \lets \Delta \lambda_0
\lets 10^{-3}.
\eeq  
In other words, if a difference of the size of the two slope parameters is 
detected at all, 
$\lambda_0^{K^0 \pi^-}$ should be slightly larger than
$\lambda_0^{K^+ \pi^0}$. 

At the Callan-Treiman point $t=\Delta_{K \pi}$, the size of the scalar 
form factor is predicted as \cite{CT66}
\beq
f_0^{K \pi}(\Delta_{K \pi}) = \frac{F_K}{F_\pi} + \Delta_{\rm CT},
\eeq
where $\Delta_{\rm CT}$ is of the order $m_u, m_d, e$.
At order $p^4$ (in the isospin limit) the correction term $\Delta_{\rm CT} 
= -3.5 \times 10^{-3}$ was calculated in \cite{gl852}. 
If isospin violation is included, it is advantageous to consider the 
quantities defined in (\ref{goodcomb0}) and (\ref{goodcombp}).
At the order $p^4$, $(m_d-m_u) p^2$, $e^2 p^2$, we find
\beq
\delta_{\rm CT}^{K^0 \pi^-} \big|_{p^4, (m_d -m_u) p^2, e^2 p^2} 
 = (1.7 \pm 0.7) \times 10^{-3}
\eeq
and 
\beq
\delta_{\rm CT}^{K^+ \pi^0} \big|_{p^4, (m_d -m_u) p^2, e^2 p^2} 
= (-10.4 \pm 0.7) \times 10^{-3} .
\eeq
In spite of the large corrections to the correction term itself, the 
Callan-Treiman relation still holds with excellent precision also if 
isospin violating contributions are taken into account. 

Corrections to $\Delta_{\rm CT}$ arising at NNLO are also (potentially) 
large. At the same time, the uncertainty of the theoretical result is 
increased by the presence of $p^6$ low-energy couplings. 
Combining the two-loop result given in \cite{Bijnens-Ghorbani07} with our 
estimate for $2 C_{12}^r + C_{34}^r$, we find (in the isospin 
limit)
\beq
\Delta_{\rm CT}  =    
\big( -7.0^{+3.5}_{-3.1} \big) \times 10^{-3}.
\eeq 
The loop contributions of order $(m_d - m_u) p^4$ were considered in 
\cite{Bijnens-Ghorbani07}. The associated counterterm contributions depend 
on partly undetermined low-energy couplings. In spite of these theoretical 
uncertainties, we expect only small corrections to the Callan-Treiman 
relation with
\beq
\big| \delta_{\rm CT}^{K^0 \pi^-} \big|, \,
\big| \delta_{\rm CT}^{K^+ \pi^0} \big| \lets 10^{-2} .
\eeq

The experimental results for the scalar slope parameter found by ISTRA+, 
KTeV and KLOE are in agreement with the predictions of the standard model.  
On the other hand, the value found by NA48 can hardly be reconciled with 
our theoretical results. Furthermore,
an isospin violation in $\Delta \lambda_0$ as it would be suggested by 
the simultaneous validity of the  results of
ISTRA+ and NA48 is definitely ruled out within the standard model.

The naive pole parametrization of the scalar form factor should be 
avoided. It contains an implicit assumption of a 
relation between slope and curvature which is not fulfilled in the 
standard model. 

At the present theoretical and experimental level of precision,
the correct treatment of electromagnetic corrections in $K_{\mu 3}$ decays 
is mandatory for the extraction of form factor parameters from 
experimental data. The appropriate procedure was described in
\cite{CKNRT02}, a more detailed presentation of the numerics is in 
preparation \cite{CGN07}. 

\medskip

\noindent
{\small {\it Acknowledgements.} 
We would like to thank Vincenzo Cirigliano, Heinrich Leutwyler and 
Emilie Passemar for informative correspondence. We are grateful to 
Gerhard Ecker for useful discussions, valuable suggestions and a 
careful reading of the manuscript.
}
\medskip

\appendix

\section{}
\label{appA}
\renewcommand{\theequation}{\Alph{section}.\arabic{equation}}
\setcounter{equation}{0}

In this section we list the coefficients $a_{PQ}(t)$,
$b_{PQ}$,
$c_{PQ}(t)$, and $d_{PQ}$ given in \cite{CKNRT02}.

\beqa 
a_{K^+ \pi^0} (t) & = & \frac{2 M_K^2 + 2 M_\pi^2 - t}{4 F_0^2} \nn  
&+& \bigg(\frac{\ve^{(2)}}{\sqrt{3}}\bigg) \frac{-2 M_K^2 + 22 
M_\pi^2 
- 9 t}{4 F_0^2} 
+ 4 \pi \alpha Z  , \nn
a_{K^0 \pi^-} (t) & = & \frac{- 2 M_K^2 - 2 M_\pi^2 + 3 t}{2 F_0^2} 
\nn 
&+& \bigg(\frac{\ve^{(2)}}{\sqrt{3}} \bigg) \frac{-2 M_K^2 + 6 
M_\pi^2 - 3 t}{2 F_0^2} 
- 16 \pi \alpha Z  , \nn
a_{K^+ \eta} (t) & = & \frac{2 M_K^2 + 2 M_\pi^2 - 3 t}{4 F_0^2} \nn 
&+& \! \! \bigg(\frac{\ve^{(2)}}{\sqrt{3}}\bigg) 
\frac{6  M_K^2 - 2M_\pi^2 - 3 t}{4 F_0^2} + 12 \pi \alpha Z  .   
\eeqa

\beqa 
b_{K^+ \pi^0} & = & - \frac{\Delta_{K \pi}}{2 F_0^2} - 
\bigg(\frac{7 \ve^{(2)}}{2 \sqrt{3}}\bigg) 
\frac{\Delta_{K \pi}}{F_0^2} 
- 4 \pi \alpha Z  , \nn
b_{K^0 \pi^-} & = & - \frac{\Delta_{K \pi}}{F_0^2} - 
\bigg(\frac{\ve^{(2)}}{ \sqrt{3}} \bigg) 
\frac{\Delta_{K \pi}}{F_0^2} 
- 8 \pi \alpha Z  , \nn
b_{K^+ \eta}  & = & -  \frac{ 3 \Delta_{K \pi}}{ 2 F_0^2} +
\bigg(\frac{\sqrt{3}   \ve^{(2)}}{2}\bigg) 
\frac{\Delta_{K \pi}}{F_0^2}
- 12 \pi \alpha Z  . \nn
&&
\eeqa

\beqa 
c_{K^+ \pi^0} (t) & = & - \frac{2 M_K^2 + 2 M_\pi^2 - 3 t}{4 F_0^2} + 
\bigg(\frac{\ve^{(2)}}{\sqrt{3}} \bigg) 
\frac{- 4 M_K^2 + 3 t}{2 
F_0^2} \nn
&-& 8 \pi \alpha Z  , \nn
c_{K^0 \pi^-} (t) & = & \frac{t}{2 F_0^2}  , \nn
c_{K^+ \eta} (t) & = &  \frac{2 M_K^2 + 2 M_\pi^2 - 3 t}{4 F_0^2} 
+ \bigg( \frac{\ve^{(2)}}{\sqrt{3}} \bigg)
 \frac{ 4 M_K^2 - 3 t}{2 F_0^2}  . \nonumber \\*
&&
\eeqa

\beqa 
d_{K^+ \pi^0}  & = & - \frac{\Delta_{K \pi}}{2 F_0^2} - 
\bigg(\frac{4  \ve^{(2)}}{\sqrt{3}}\bigg)  
\frac{\Delta_{K \pi}}{F_0^2} 
+ 4 \pi \alpha Z  , \nn
d_{K^0 \pi^-}  & = & - \frac{\Delta_{K \pi}}{F_0^2} - 
\bigg(\frac{2  \ve^{(2)}}{ \sqrt{3}}\bigg) 
\frac{\Delta_{K \pi}}{F_0^2} 
+ 8 \pi \alpha Z  , \nn
d_{K^+ \eta}  & = & -  \frac{ 3 \Delta_{K \pi}}{ 2 F_0^2}  
+ 12 \pi \alpha Z  . 
\eeqa

\end{document}